\documentclass[useAMS,usenatbib]{mn2e}

\usepackage{graphics}
\usepackage{amsmath}
\usepackage{txfonts}
\usepackage{rotating}
\usepackage{textcomp}
\usepackage{amssymb}

\usepackage{fixltx2e}
\usepackage{url}
\usepackage{natbib}

\usepackage{booktabs}
\usepackage[normalem]{ulem}
\usepackage{colortbl, xcolor}
\usepackage{multirow}
\usepackage{threeparttable}

\def\black{\color{black}}

\def\kms{km~s$^{-1}$}
\def\degree{$^{\rm o}$}
\def\kmskpc{km~s$^{-1}$~kpc$^{-1}$}

\def\Omegap{\mbox{$\Omega_{\rm p}$}}
\def\cp{\mbox{P$_{\rm c, \emph m}$ }}
\def\sp{\mbox{P$_{\rm s, \emph m}$ }}
\def\Epic{\mbox{\textsc{Epic5}}}
\def\msun{\mbox{M$_{\odot}$}}

\defcitealias{Wada1994}{W94}

\title[Gas orbits in a rotating galactic potential]{Analytic gas orbits in an arbitrary rotating galactic potential using the linear epicyclic approximation}
\author[N. Pi\~nol-Ferrer, P. O. Lindblad \& K. Fathi]{N. Pi\~nol-Ferrer$^{1}$\thanks{E-mail: npi@astro.su.se}, 
P. O. Lindblad$^{1}$, K. Fathi$^{1}$\\
$^1$Stockholm Observatory, Department of Astronomy, Stockholm University, AlbaNova Center, 106 91 Stockholm, Sweden\\
}

\pagerange{\pageref{firstpage}--\pageref{lastpage}} \pubyear{2011}

\begin{document}

\label{firstpage}

\maketitle

\begin{abstract}
\indent
A code, \Epic, has been developed which computes, in the two-dimensional case, 
the initially circular orbits of guiding centra in an arbitrary 
axisymmetric potential with 
an arbitrary, weak perturbing potential in solid body rotation. This 
perturbing potential is given by its Fourier expansion. The analytic solution 
solves the linear epicyclic approximation of the equations of motion. 
To simulate the motion of interstellar matter and to damp the Lindblad 
resonances, we have in these equations introduced a friction which is proportional to the deviation from circular velocity. The 
corotation resonance is also damped by a special parameter. The program 
produces, in just a few seconds, orbital and density maps, as well as line 
of sight velocity maps for a chosen orientation of the galaxy. 

     We test \Epic~ by comparing its results with previous simulations and 
observations from the literature, which gives satisfactory agreement. The aim 
is that this program should be a useful complement to elaborate numerical 
simulations. Particularly so are its abilities to quickly explore the parameter 
space, to construct artificial galaxies, and to isolate various single agents 
important for developing structure of interstellar matter in disc galaxies.
\end{abstract}

\black

\section{Introduction}

Since the emergence of high spatial resolution two-dimensional velocity field for galaxies, the interpretation of observed velocity fields has become one of the key concerns of the extragalactic scientific community (e.g., \citealt{Boulesteix1987}; \citealt{Adam1989}; \citealt{Bacon1992}; \citealt{Bacon2001}; \citealt{Allington1997}; \citealt{Hernandez2003}). Of particular importance is the class of galaxies which host bars and/or spiral arms, as they are estimated to be the most abundant types of galaxies in the local Universe \citep{Loveday1996}.

The theoretical foundation for understanding the kinematics and dynamics of spiral galaxies has been worked out over the last century. 
Bertil \citet{Lindblad1963,Lindblad1964} conceived a picture of circulation of stars between the spiral arms of a quasi-steady rotating spiral potential. Simultaneously, \citet{LinShu1964} presented their density wave theory approaching the problem from a different theoretical point of view. Subsequently, \citet{Shu1973} and \citet{Roberts1979} derived solutions for the circulation of the interstellar medium through such a spiral potential and predicted large scale galactic shocks along the spiral arms. These processes make bars and/or spiral arms potential actors to redistribute angular momentum which will lead to dramatic effects such as intense star formation, build up of bulges, or onset of nuclear activity. In turn, the host galaxies may undergo dramatic secular evolution on time-scales compared to the galaxy dynamical time (see \citealt{Kormendy2004} for a comprehensive discussion).

Beginning with \citet{Holmberg1941} in the pre-computer era and in the 1950s with electronic computers (e.g. P.O. \citealt{Lindblad1960}), a large number of simulations of development and evolution of structure in galaxies have been made with a variety of computer codes. These studies have often aimed at establishing the details of the flow of gas in and around bars (eg. \citealt{Athanassoula1992b}), as well as the mechanisms which trigger starbursts and nuclear activity in the centers of galaxies (e.g., \citealt{Shlosman1989}). However, they often involve extensive computational power leading to relatively long computing times. 


Several successful attempts have been carried out in order to deliver analytic solutions for observable parameters for some of the effects involved in the above questions (Sakhibov \& Smirnov 1989; \citealt{Canzian1993}; \citealt{Wada1994}; \citealt{Lindblad1994}; \citealt{Schoenmakers1997}; \citealt{Wong2004}; \citealt{Fathi2005}; \citealt{Byrd2006}; \citealt{vandeVenfathi2010}), though they have all exclusively treated one specific morphological feature in each analytic model. While a linear analysis can be a guide for a physical understanding of complicated numerical results from simulations, in these analytic studies, bars and spiral arms have been accounted for separately, and interpretations have been based on marginalizing the effect of the other features. A notable difference between these results and numerical simulations is that the stepwise integration in a detailed simulation requires appreciable computing times, while for analytical calculations the computing time is negligible.


In this paper, we present an analytic solution within the epicyclic approximation, in which we introduce an arbitrary gravitational potential 
and a damping coefficient for adequate appearance of the corotation resonance. Our solution has been coded as a development of the program \textsc{Epic} \citep{Lindblad1994} which was based on the first order epicyclic approximation. The current code, \Epic, derives the response of interstellar matter, originally in circular orbits, to the gravitational potential 
in a matter of few seconds, which makes it an efficient code for surveying a large parameter space. In section \ref{sec:method}, we present the analytical solution for the epicyclic theory with a generic potential, and we describe our code \Epic. We compare results generated by \Epic~ with \citet{Wada1994} in section \ref{sec:models}, and in section \ref{sec:new}, we demonstrate a standard case of a barred galaxy. In section \ref{sec:1365} we compare results obtained by \Epic~with observations and simulations made for the galaxy NGC~1365. Finally, we conclude in section \ref{sec:conclusions}.

\section{Method}
\label{sec:method}


The epicyclic description of nearly circular stellar orbits in a circularly symmetric galaxy was developed by B. \citet{Lindblad1927}  in order to theoretically explain the observed velocity ellipsoid in the Milky Way galaxy. Later on, B. \citet{Lindblad1958}  introduced a rotating perturbing potential in the theory and pointed out the resonances that carry his name.
 P.O. Lindblad \& P.A.B. Lindblad (1994), as well as \citet{Wada1994}, introduced gas dynamical friction in the epicyclic approximation. This form of damping the motions would make the first order approximation valid also over the Lindblad resonances, but still not over the corotation resonance. 

The potential we are considering in our problem is:
\begin{equation}
 \Phi (r,\theta) = \Phi_0 (r) + \Phi_1 (r,\theta)
\label{eq:pot}
\end{equation}
\noindent
where ($r,\theta)$ are the polar coordinates in a frame co-rotating with the potential at the pattern speed frequency \Omegap. $\Phi_0(r)$ is the axisymmetric potential, and $\Phi_1 (r,\theta)$ is an arbitrary perturbing potential developed as a Fourier series

\begin{equation}
\centering
\Phi_1 (r,\theta) = - \sum_{m=1}^{\rm n} \left[ \rm P_{ \emph \rm c,\emph m} (\emph r) \cos\left( \emph m\,\theta\right) + P_{\rm s,\emph m}(\emph r) \sin\left(\emph m\,\rm \theta\right)\right] 
\label{eq:fourierpot}
\end{equation}

\noindent
or

\begin{equation}
\centering
\Phi(r,\theta)=\Phi_0(r)-\sum_{\emph m=1}^{\rm n} \Psi_{\emph m}(r) \cos\emph m\left(\theta-\vartheta_{\emph m}(r)\right) 
\end{equation}

\noindent
where

\begin{equation}
\centering
\Psi_{\emph m}^2(r)=\rm P_{ \emph \rm c,\emph m}^2 + P_{\rm s,\emph m}^2 \,\,\,\,\,\,\,\,\text{     and     }\,\,\,\,\,\,\,\,
\tan(\emph m \vartheta_{\emph m})=\rm P_{ \emph \rm c,\emph m} / P_{\rm s,\emph m} 
\end{equation}

We consider small deviations, $\xi$ and $\eta$, from circular motion following the usual notation, where 
\begin{align}
r&=r_0+\xi \\
\theta&=\theta_0 + (\Omega-\Omegap) \, t + \frac{1}{r_0} \eta
\label{eq:coor}
\end{align}

\noindent
and $\Omega(r_0)$ is the angular frequency of circular motion. 
In addition, we introduce a frictional force proportional to the velocity deviation from circular motion with the coefficient $-2\lambda$. As \citet{Wada1994} points out, this is analogous to the Stokes' formula \citep[eg.][p. 365]{Yih1977} where the drag on a sphere moving slowly in a viscous fluid is proportional to the first power of the velocity. The equations of motion, as derived in appendix \ref{ap:theory}, can then be written as:

\begin{align}
\centering
\ddot{\xi} + 2 \lambda \dot{\xi} - 2\Omega \dot{\eta} - 4 \Omega A \xi  &= - \frac{\partial \Phi_1}{\partial r}  =\nonumber\\
=&\sum_{m=1}^{n} \big[ C_m \cos m(\theta - \vartheta_m) + E_m \sin m(\theta - \vartheta_m)\big] \label{eq:eqofmotionfric1}\\
\ddot{\eta} + 2 \Omega \dot{\xi} + 2 \lambda \dot{\eta} + 4 \lambda A \xi &= - \frac{1}{r}\frac{\partial \Phi_1}{\partial \theta} =
- \sum_{m=1}^{n} D_m \sin m(\theta - \vartheta_m) \label{eq:eqofmotionfric2}
\end{align}

\noindent
where  
\begin{equation}
C_m=\frac{d\Psi_m}{dr}; \,\,\,\,\,\,D_m=m\frac{\Psi_m}{r}; \,\,\,\,\,\, E_m=m\Psi_m\frac{d\vartheta_m}{dr} \nonumber
\end{equation}

The solution for the motion of the guiding center will then be:
\begin{align}
\xi = \sum_{m=1}^{n} \big[ d_m \cos m(\theta-\vartheta_m) + e_m \sin m(\theta -\vartheta_m) \big] \label{eq:xisol}\\
\eta= \sum_{m=1}^{n} \big[ g_m \sin m(\theta-\vartheta_m) + f_m \cos m(\theta -\vartheta_m) \big]\label{eq:etasol} 
\end{align}

\noindent
where the amplitudes $d_m$, $e_m$, $g_m$ and $f_m$ are given in appendix \ref{ap:theory}.

\subsection{The code: \Epic}
\label{subsec:epic}

The code \Epic ~is an extension of the code developed by \citet{Lindblad1994} and computes the analytic solution (\ref{eq:xisol}) and (\ref{eq:etasol}),
as well as its corresponding density and velocity maps, generated by the arbitrary potential.
\Epic~derives the axisymmetric potential, $\Phi_0$, from a given rotational velocity curve and the perturbing potential, $\Phi_1$, is introduced by its Fourier decomposition, \cp and \sp (eq. \ref{eq:fourierpot}).  

In addition to the velocity curve and the parameters of the potential, another free input parameter
is a constant pattern speed of the potential, \Omegap. The damping coefficient of the frictional force, $\lambda$, which was introduced in the analysis to simulate orbits of interstellar matter, is assumed to have a linear tendency with radius. This linearity is determined by the given values of $\lambda$ at the outer inner Lindblad resonance (oILR) and at the outer Lindblad resonance (OLR) of the system (it is taken at 0~kpc if no ILR is present and if no OLR is present, it is taken at $r_{max}$ in the rotation curve). These two values are also given as input parameters in \Epic. 
This definition of the damping coefficient, $\lambda$, that allows it to vary along the radius, is desirable since it should depend on the gas density.

As can be seen from eqs. \ref{eq:d}-\ref{eq:f}, the introduction of the friction coefficient $\lambda$ has damped the amplitudes and eliminated the singularities at the Lindblad resonances where $\kappa^2 - \omega_m^2 = 0$. However, the singularity at corotation, $\omega_m=0$, remains. As shown by \citet[Ch 3.3.3(b)]{BinneyTremaine2008}, the stellar motions close to corotation turn into pendulum like oscillations around the Lagrange equilibrium points $L_4$ and $L_5$. With increasing distance from corotation, the angular amplitude increases until the orbit reaches the equilibrium points $L_1$ and $L_2$ and flips into a circumcentral orbit (see Figs. 16 and 26 in P.A.B. \citealt[hereafter LLA96]{Lindblad1996}). In the simulations of LLA96 the points $L_4$ and $L_5$ represent density minima and $L_1$ and $L_2$ density maxima along the corotation radius.

Actually, our equations of motion (\ref{eq:xisol}) and (\ref{eq:etasol}) are no longer valid around corotation because when inserting $\theta$ from eq. (\ref{eq:coor}) into the differential equations (\ref{eq:eqofmotionfric1}) 
and (\ref{eq:eqofmotionfric2}) we have assumed $\eta/r_0 << (\Omega-\Omega_p)$ which is not true close 
to corotation where $\Omega-\Omega_p$ approaches zero. We then see from eq. (\ref{eq:eqofmotionfric2}) 
that for small $\xi$, $\dot{\xi}$ and $\dot{\eta}$ we get a pendulum like oscillation in 
$\eta$, and we should not get a linearized equation. 

\begin{figure}
   \centering
\includegraphics[width=0.48\textwidth, trim=19mm 13mm 1mm 0mm]{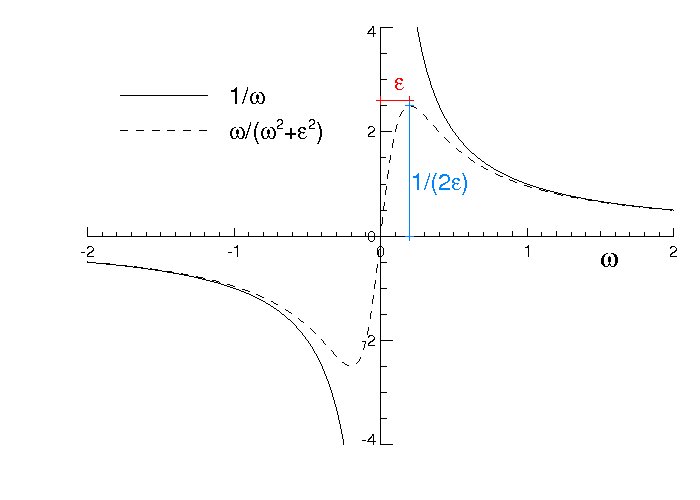}
\caption{Damping of the amplitudes $\xi$ and $\eta$ at corotation. \emph{Solid line}: Variation of the amplitudes with $\omega$ according to the first order epicyclic theory with $\varepsilon=0$. \emph{Dashed line}: Variation of the amplitudes introduced in \Epic.}
\label{fig:epsplot}
\end{figure}

    P.O. \citet{Lindblad1960} has pointed out that a time dependent potential 
can damp all resonances, including the corotation resonance. 
For a potential that varies as $e^{\gamma t}$, besides the twisting of the orbits around corotation, $\omega_m$ in the 
denominator of eqs. (\ref{eq:d}) to (\ref{eq:f}) is replaced by 
$\sqrt(\omega_m^2+\gamma^2)$. In similarity to this, in \Epic~the corotation singularity is avoided by replacing the factor $\omega_m$ in the denominator of eqs. (\ref{eq:d})-(\ref{eq:f}) by $\omega_{\epsilon}$, where

\begin{align}
\frac{1}{\omega_{\varepsilon}}=\frac{\omega_m}{\omega_m^2+\varepsilon_m^2}
\end{align}

\noindent
where $\varepsilon_m=m\varepsilon$ and $\varepsilon$ is an additional input parameter in the code. As illustrated in Fig. \ref{fig:epsplot}, the amplitudes of the motion of the guiding center, growing large and of opposite sign on both sides of corotation, are smoothed and brought to zero at the exact resonance distance. $1/\omega_{\varepsilon}$ reaches a maximum of $1/2\varepsilon_m$ at the distance $\omega_m=\varepsilon_m$ from corotation. This means that the stellar elliptical orbits, trapped around corotation, in our case collapses to circular rotation at the corotation radius. 
This may not be an entirely inappropriate approximation to the solution in the corotation case.

The derivation of the density is made 
applying the continuity equation. Using the equation (F-5) in \citet{BinneyTremaine2008}, in our notation the ratio between the perturbed and unperturbed surface densities will be

\begin{equation}
\frac{\rho}{\rho_0} = 1 - \xi- \frac{1}{r}\frac{\partial \eta}{\partial \theta} - \frac{\partial \xi}{ \partial r}
\label{eq:density}
\end{equation}

\noindent
where we have assumed the unperturbed surface density $\rho_0$ to have a flat distribution.

\Epic~generates a number of supporting plots and diagrams like the rotation curve, the circular frequencies showing the resonances, $\omega$ vs $\kappa$, the perturbing potential, the total potential, radial and tangential perturbing forces, the phase variation of the perturbing potential, orbits, densities, as well as line of sight velocity fields for various position angles of the line of nodes.

The resolution in $r$ required by \Epic~in the tabulated values of the Fourier components of the perturbing potential is related to the increase of the phase shifts $\vartheta_m$ with $r$. It is convenient to express this relation in terms of the wavelength of the mode $m=2$ in the spiral structure. If $\lambda_m$ is the wavelength of mode $m$, we have approximately for a tightly wound spiral
%
\begin{align}
\frac{\rm d\vartheta_{\emph m}}{\rm d \emph r} = \frac{2\pi}{\emph m \lambda_m} = \frac{\pi}{\lambda_2}
\label{eq:k}
\end{align}
Our requirement for \Epic~to interpret the potential tables $\rm P_{ \emph \rm c,\emph m}(r)$, $\rm P_{\rm s,\emph m}(r)$ is $\Delta(m\vartheta_m)<\pi/2$ or
\begin{align}
\Delta r<\frac{\lambda_2}{  2\,n\, }
\label{eq:condition}
\end{align} 
%
where $\Delta r$ is the radial increment in the potential table and $n$ is the maximum value of $m$.

Finally, we can choose between counter-clock and clock wise rotation for an easier comparison with observed cases. 

\section{\Epic~vs Wada's (1994) model}
\label{sec:models}

\citet[hereafter W94]{Wada1994} independently provided an analytical model, similar to \textsc{Epic}, representing the behavior of a non-self-gravitating gas in a rotating potential with a weak bar-like distortion. Like the original version of \textsc{Epic} \citep{Lindblad1994} there was no solution for the corotation region. 


In the model Wada used the Toomre potential for the axisymmetric potential, and for the perturbation the potential given by \citet{Sanders1977}

\begin{equation}
\Phi(R,\psi)=\Phi_0(R) ( 1 + \Phi_{\rm b}(R ) \cos{2\psi})
\end{equation}

\begin{equation}
\Phi_0 = - \sqrt{\frac{27}{4}} \frac{a \,v^2_{\rm max}}{\sqrt{R^2+a^2}} \;\;\;\;\; \& \;\;\;\;\;
\Phi_{\rm b} = \varepsilon_0 \, a \frac{R^2}{(R^2+a^2)^{1.5}}
\label{eq:wadapot}
\end{equation}

In these equations, $v_{\rm max}$ is the maximum rotational velocity, $a$ is the core radius and $\varepsilon_0$ is a parameter which represents the strength of the bar potential. 


We test our code by using the same potential, circular frequency and parameters as W94. W94, however, chooses the friction to be proportional to the velocity in the radial direction only.


\citetalias{Wada1994} considers a core radius, $a$, equal to 1 and a maximum rotational velocity, $v_{\rm max}$, equal to $(4/27)^{1/4}$, simplifying the expressions of $\Phi_0$ and $\Phi_{\rm b}$. W94 assumed a weak bar strength, $\varepsilon_0$, equal to 0.05 and a pattern speed equal to 0.1. 

\begin{figure}
   \centering
\includegraphics[width=0.78\linewidth, trim=0mm 0mm 3mm 0mm, angle=-90]{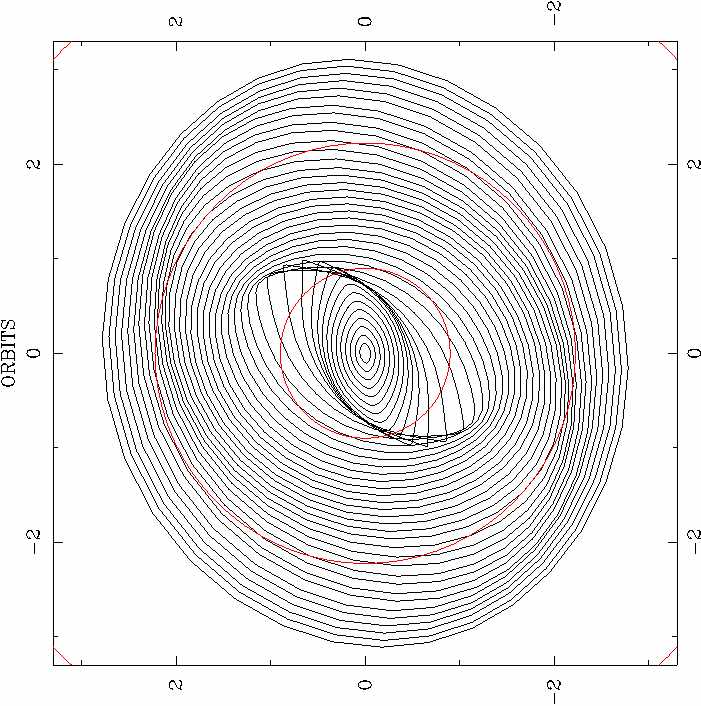}\\ \vspace{0.4cm}
\includegraphics[width=0.78\linewidth, trim=0mm 0mm 3mm 0mm, angle=-90]{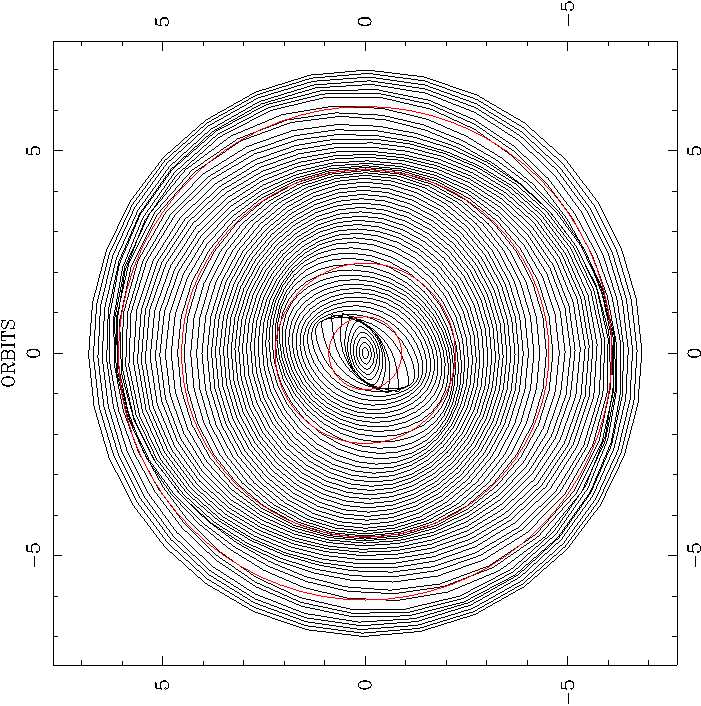} \\ \vspace{0.4cm}
\includegraphics[width=0.98\linewidth, trim=0mm 0mm 3mm 0mm, angle=-90]{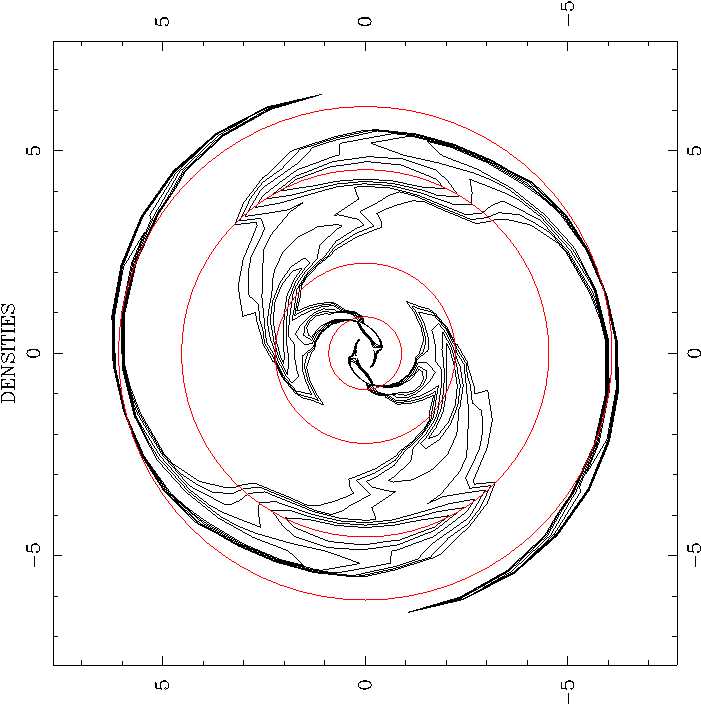} 
\caption{Results from \Epic~for the W94 case. \emph{Upper graph:} Orbits generated by \Epic~inside corotation. \emph{Middle graph:} Orbits generated by \Epic~including the outer Lindblad resonance (OLR). \emph{Lower graph:} Density contours given by \Epic~according to eq. (\ref{eq:density}). Contours at 1.01, 1.05, 1.1, 1.2, 1.3 and 1.4. Red circles show the positions of resonances iILR, oILR, CR and OLR. The rotation is counter clockwise.}
\label{fig:wada94mod1}
\end{figure}

We use in this section the same potentials and derive the input needed for \Epic~from them (the rotational velocity curve and the Fourier components of the perturbed potential). We have also used the same pattern speed of 0.1, a friction coefficient of 0.02 at the oILR and 0.01 at the OLR and a corotation softening coefficient of 0.02. We have located the bar horizontally, equally to its position in W94, and consider a counter-clockwise rotation.

Using these input parameters for our code, we have derived the resonance radia at the same radia as W94: inner/outer Lindblad resonance (iILR and oILR) at 0.90 and 2.22 respectively, corotation (CR) at 4.53 and outer Lindblad resonance (OLR) at 6.09 (see Fig. 1 in W94). We present the orbits generated by \Epic~in the upper panel of Fig. \ref{fig:wada94mod1}, corresponding to the ones presented in Fig.~4 in W94. The leading arms around the iILR and trailing arms around oILR are well presented in both cases. \Epic~damps as well the orbits around corotation, allowing us to present orbits until radia further than CR (see middle panel in Fig. \ref{fig:wada94mod1}).

\Epic~estimates the density map generated by the input potential in the lower panel in Fig. \ref{fig:wada94mod1}, showing the continuation of the spiral arms out to the OLR. 


\section{\Epic: a demonstration}
\label{sec:new}

\begin{figure}
\centering
\includegraphics[width=0.36\textwidth, trim=0mm 0mm 0mm 0mm, angle=-90]{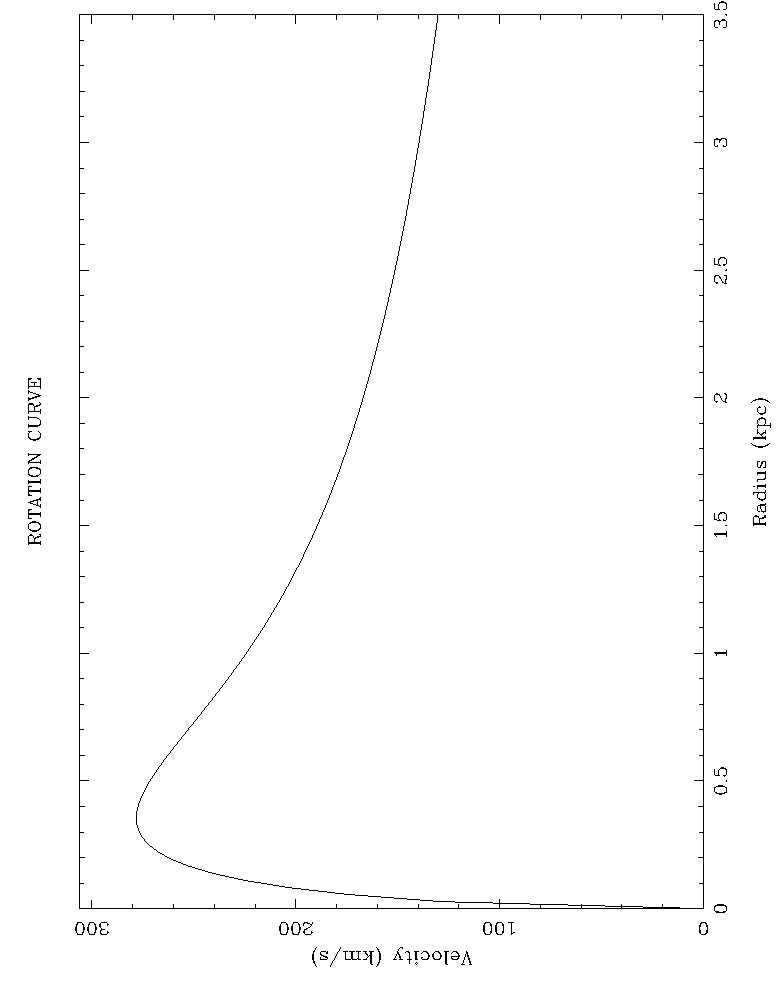} 
\caption{Rotation curve given by two superimposed exponential disks with masses of $11.5 \times 10^9$~\msun~and $2.87 \times 10^9$~\msun~and with scale lengths of 0.194~kpc and 1.15~kpc respectively.}
\label{fig:rotdemos}
\end{figure}

\begin{figure*}
\centering
\includegraphics[width=0.48\linewidth, trim=0mm 0mm 0mm 0mm, angle=-90]{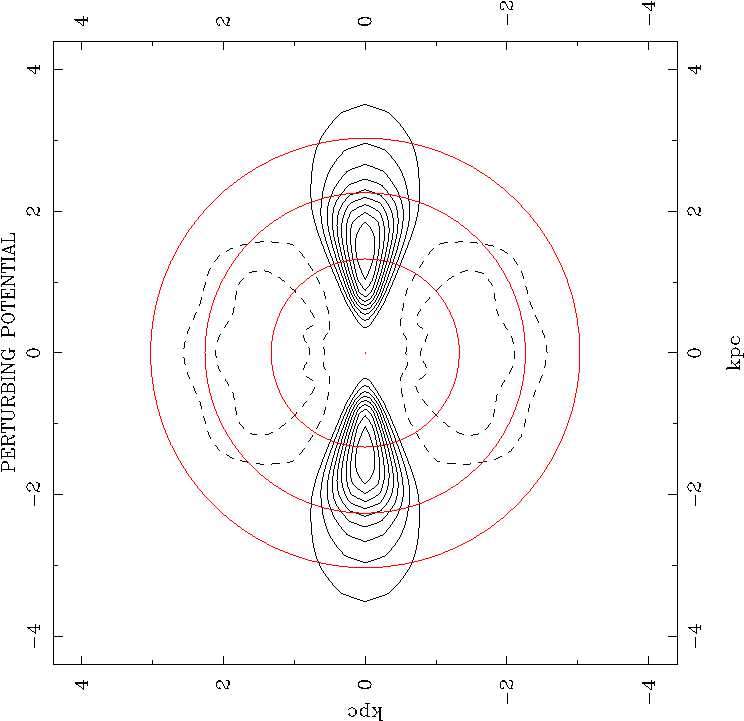} \label{fig:modpo_pota}
\includegraphics[width=0.48\linewidth, trim=0mm 0mm 0mm 0mm, angle=-90]{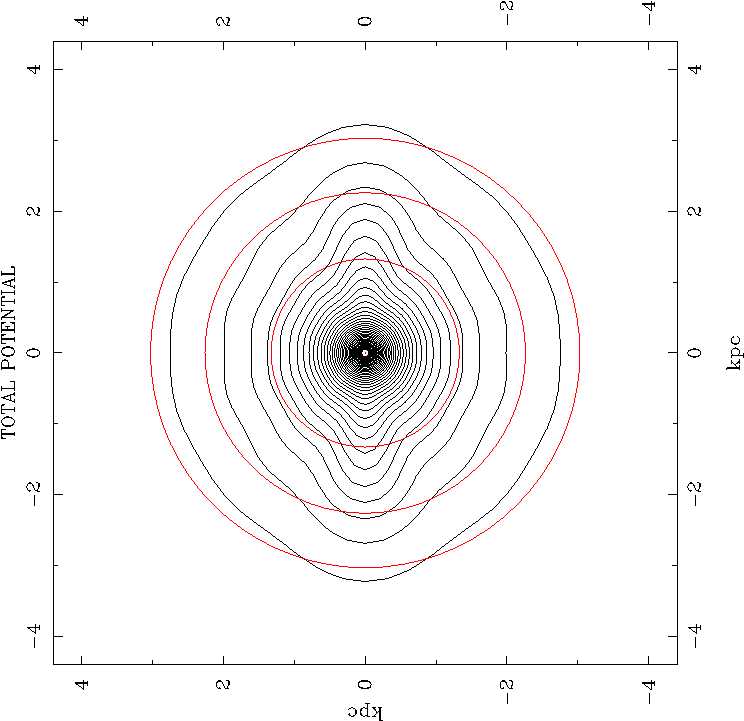}\label{fig:modpo_potb}
\caption{Bar potential. \emph{Left graph:} Perturbing potential. Dashed lines show the positive contribution and the solid lines the negative contribution. The red circles mark the oILR, corotation and OLR for a pattern angular velocity of 70~\kmskpc. \emph{Right graph:} Total potential.}
\label{fig:modpo_potb}
\end{figure*}

We demonstrate in this section a case to illustrate what \Epic~is able to produce. The rotation curve in this case, shown in Fig. \ref{fig:rotdemos}, is that given by two superimposed exponential disks with masses of $11.5 \times 10^9$~\msun~and $2.87 \times 10^9$~\msun~and with scale lengths of 0.194~kpc and 1.15~kpc respectively (reasonable values for the inner part of barred galaxies, \citealt{Lindblad2010}). To this, is added an ad hoc perturbing bar potential, described by its Fourier components of cosine of 2, 4, 6, 8 and 10 times $\theta$ for each value of the radius, as seen in Fig. \ref{fig:modpo_pota} (left graph). This results in the total potential shown in Fig. \ref{fig:modpo_potb} (right graph). The relative bar strength $|\Phi_1|_{\rm min}/|\Phi_0|_{\rm min}$ is equal to 0.026.

\begin{figure}
\centering
\includegraphics[width=0.7\linewidth, trim=0mm 0mm 0mm 0mm, angle=-90]{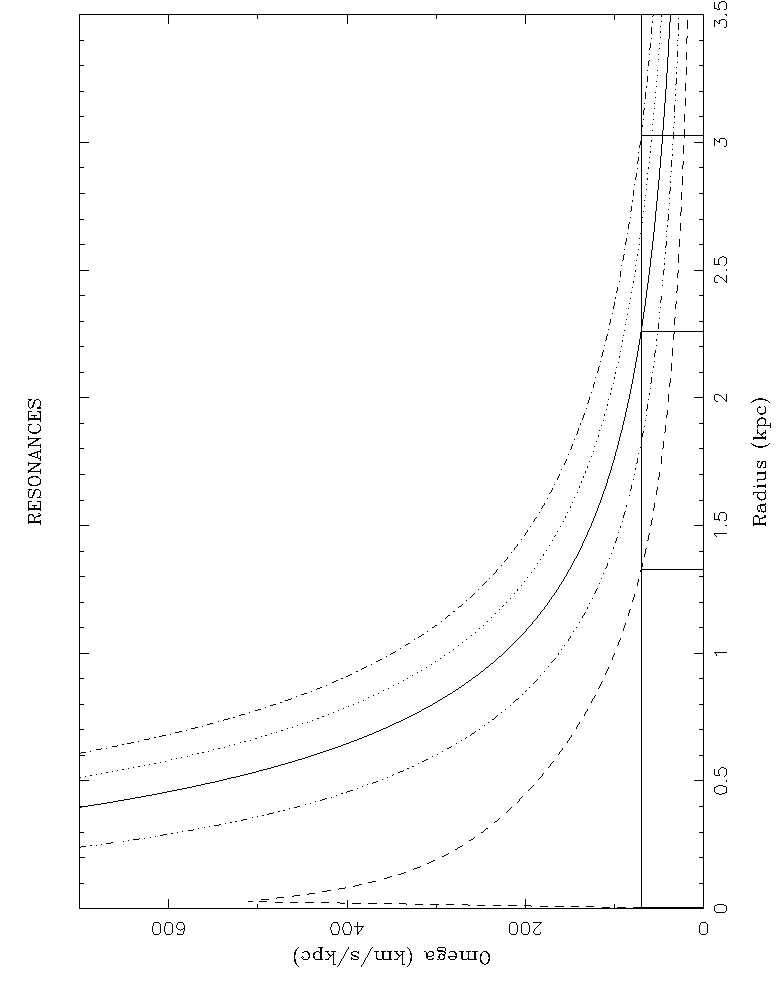}
\caption{Angular velocity, $\Omega$, against radius derived from the rotation curve (solid line). The dashed line shows $\Omega-\kappa/2$, the dotted-dashed line shows $\Omega+\kappa/2$, the dotted and dashed-triple dotted lines show $\Omega\pm\kappa/4$. The pattern speed of 70~\kmskpc~is shown by the horizontal solid line, and the resonances are indicated by the vertical lines.}
\label{fig:modpo_ome}
\end{figure}

Fig. \ref{fig:modpo_ome} shows the variations of $\Omega$, $\Omega\pm \kappa/2$ and $\Omega\pm \kappa/4$ with radius. Our assumed pattern velocity of 70~\kmskpc, which places the oILR close to the minimum of the perturbing potential, is shown as a straight line in the figure, and the resonances are shown by the vertical lines. The iILR lies very close to the center, oILR at 1.33~kpc, CR at 2.26~kpc, and OLR at 3.03~kpc.

\begin{figure}
\centering
\includegraphics[width=0.9\linewidth, trim=0mm 0mm 0mm 0mm, angle=-90]{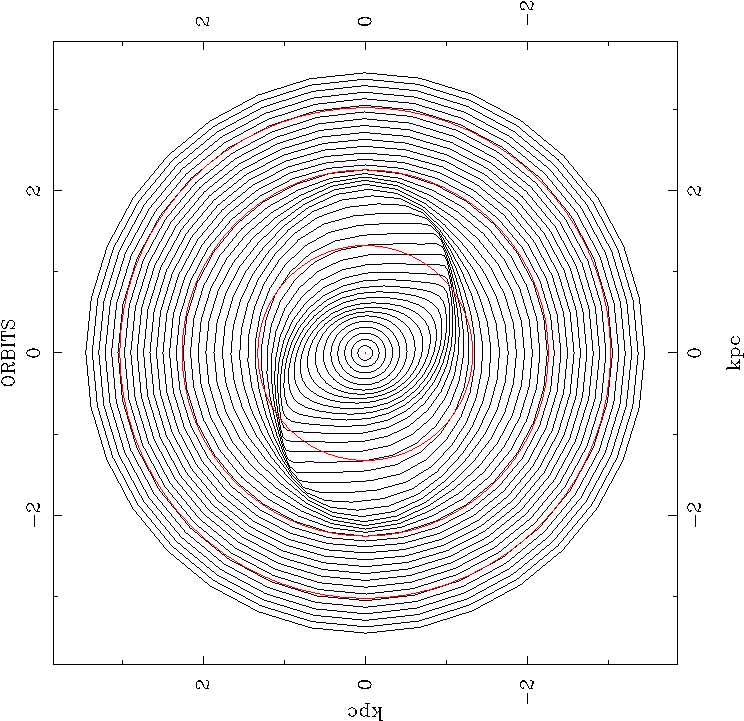}
\caption{Orbits generated with \Epic. The red circles mark the oILR, corotation and OLR. The rotation is clockwise.}
\label{fig:modpo_orb}
\end{figure}

\begin{figure*}
\centering
\includegraphics[width=0.48\linewidth, trim=0mm 0mm 0mm 0mm, angle=-90]{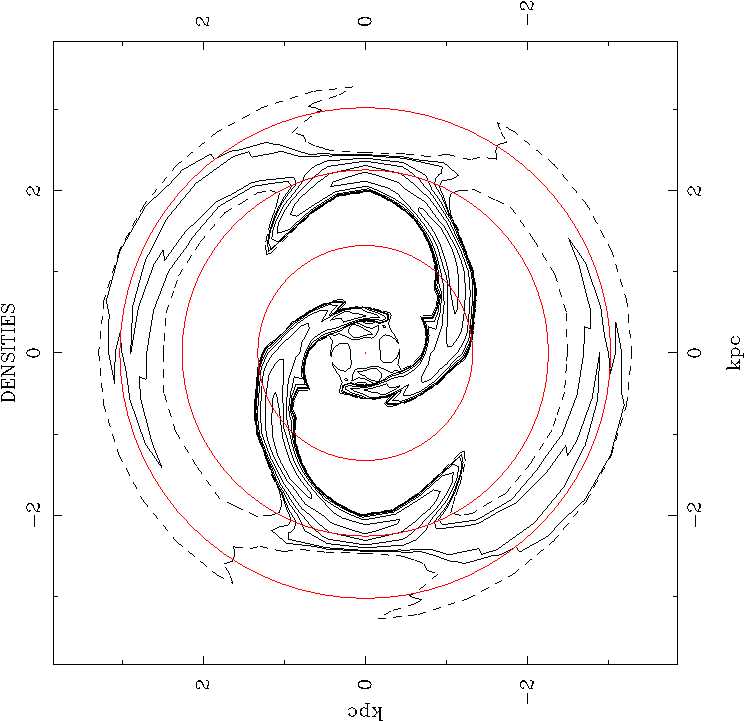}
\includegraphics[width=0.48\linewidth, trim=0mm 0mm 0mm 0mm, angle=-90]{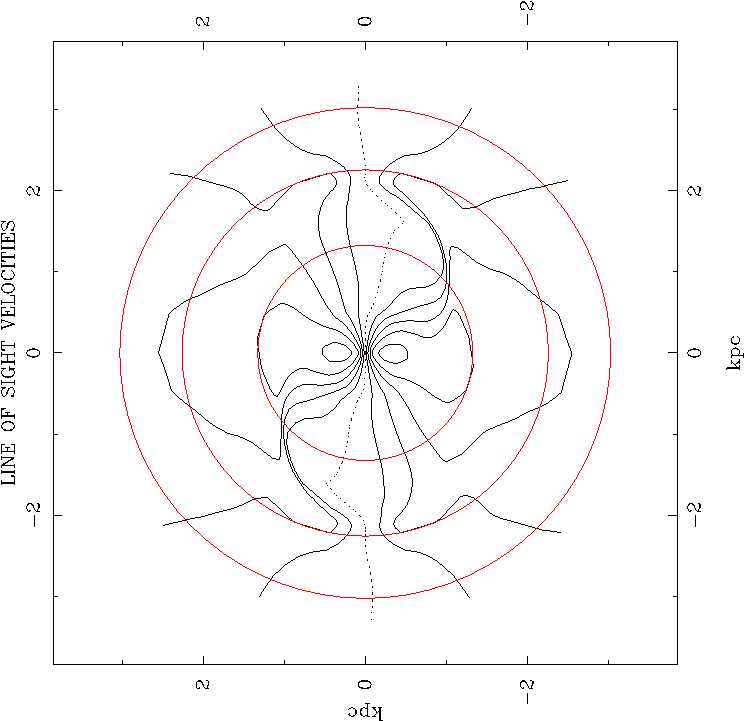}
\caption{\emph{Left graph:} Density contour $\rho/\rho_1$ in the plane of the galaxy. Contours at 0.99, 1.01, 1.02, 1.05, 1.1, 1.2, 1.4 and 1.7. \emph{Right graph:} Velocity field in the plane of the galaxy with a position angle of 0\degree~and an increment of velocity between contours of 50~\kms. The direction of rotation is clockwise.}
\label{fig:modpo_denvel}
\end{figure*}

Computing the orbits, we have assumed the damping coefficients constant at $\lambda=60$~\kmskpc~and  $\varepsilon=60$~\kmskpc. The resulting orbits are seen in Fig. \ref{fig:modpo_orb}. The damping coefficients have been adjusted such that the orbits, presenting
a laminar flow pattern, do not cross. This is also a condition for \Epic~to 
be able to produce meaningful density and velocity maps. This choice means 
that these coefficients and the strength of the bar are correlated. The densities, as given by \Epic~are shown in Fig. \ref{fig:modpo_denvel} (left graph), and the line of sight velocities in the plane of the galaxy as seen with a position angle of the line of nodes of 0\degree~ is given in Fig. \ref{fig:modpo_denvel} (right graph). 

We see how the orbits well inside the oILR are orientated perpendicular to the bar but twisting against the direction of rotation. At the oILR they are tilted about 45\degree~against the position of the bar, and close to corotation they are elongated along the bar.  The orbits show sharp kinks on the leading side of the bar. This is connected to large velocity jumps at the corresponding positions and strong enhancements of the intensities parallel to the bar on the leading side from the ILR to CR. 
Such dust lanes along the leading side of the bar are well-known common features in barred galaxies (e.g. \citealt{Athanassoula1992b}). 
From corotation the response splits and a trailing spiral arm continues out to OLR. The very faintest density contours close an ellipse between CR and OLR elongated perpendicular to the bar.

\begin{figure*}
\centering
\includegraphics[width=0.48\linewidth, trim=0mm 0mm 0mm 0mm, angle=-90]{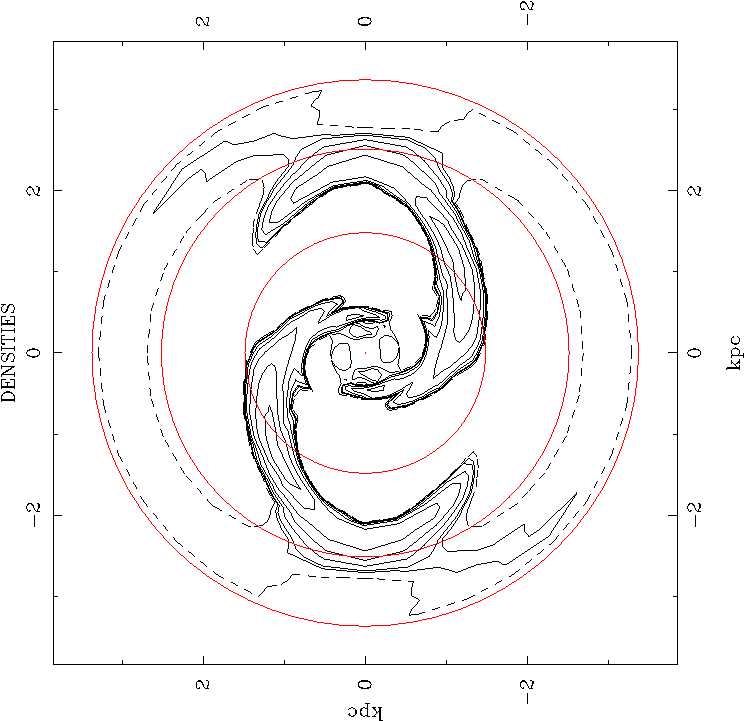} \label{fig:den60}
\includegraphics[width=0.48\linewidth, trim=0mm 0mm 0mm 0mm, angle=-90]{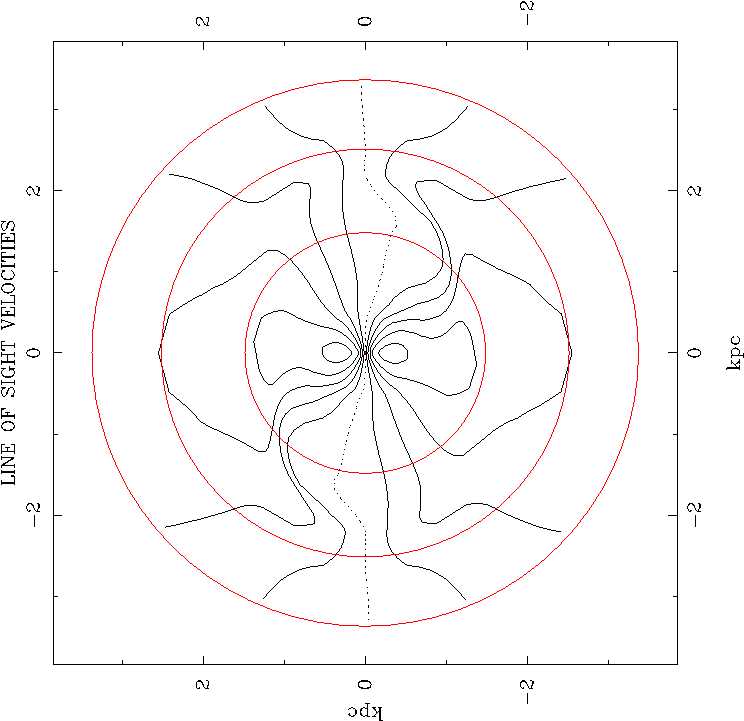}\label{fig:vel60}\\ \vspace{0.1cm}
\includegraphics[width=0.48\linewidth, trim=0mm 0mm 0mm 0mm, angle=-90]{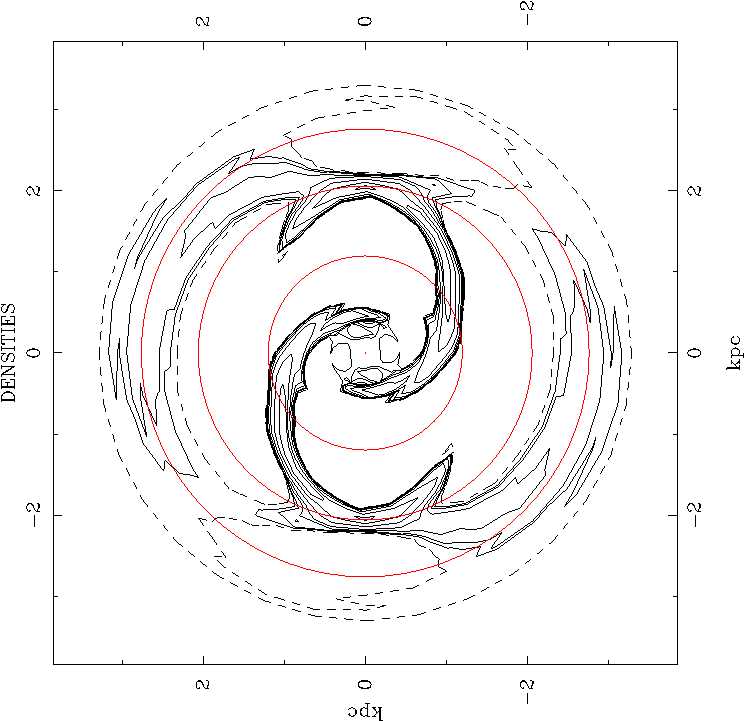} \label{fig:den80}
\includegraphics[width=0.48\linewidth, trim=0mm 0mm 0mm 0mm, angle=-90]{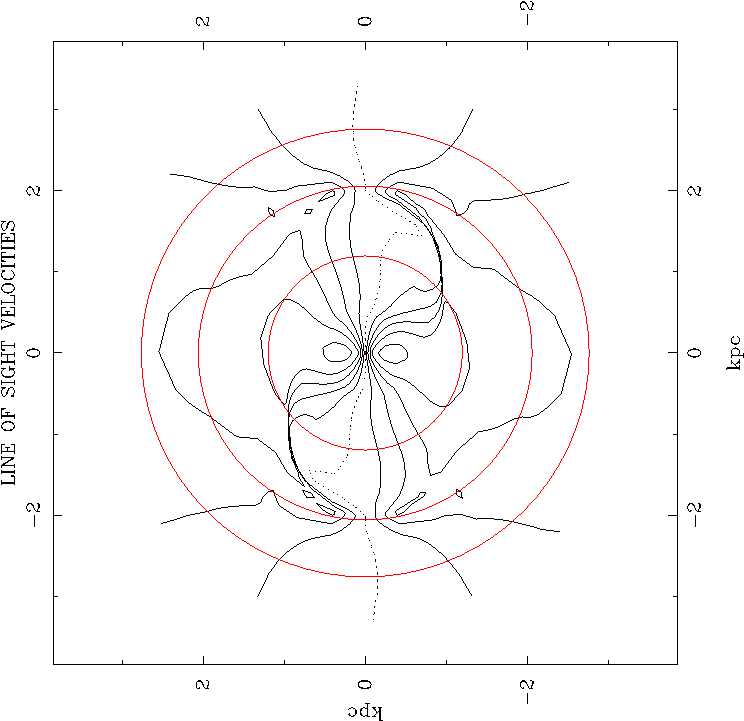}\label{fig:vel80}
\caption{\emph{Upper and lower left graph:} Density contour $\rho/\rho_1$ in the plane of the galaxy, for $\Omega_p$ equal to 60~\kmskpc and 80~\kmskpc, respectively. Contours at 0.99, 1.01, 1.02, 1.05, 1.1, 1.2, 1.4 and 1.7. \emph{Upper and lower right graph:} Velocity field in the plane of the galaxy with a position angle of 0\degree~and an increment of velocity between contours of 50~\kms. Upper graph for a $\Omega_p$ equal to 60~\kmskpc~and lower for 80~\kmskpc. The direction of rotation is clockwise.}
\label{fig:case6080}
\end{figure*}

Using the same case, we have varied $\Omega_p$ to observe changes in the response. The density and velocity maps generated by \Epic~for $\Omega_p$ equal to 60~\kmskpc~ and to 80~\kmskpc~ are shown in Fig.~\ref{fig:case6080}. We have used a constant damping coefficient $\lambda$ and $\varepsilon$ equal to 60~\kmskpc~for the slow and fast potential. The trailing arms extending from CR and going towards OLR are shorter and fainter as the potential gets slower. Moreover, a faster bar generates more prominent, and also more tightly wound, dust lanes inside CR, with larger velocity jumps, as shown in Fig. \ref{fig:case6080}, right graphs. When the pattern speed gets larger, the resonance distances decrease. As seen from Fig.~\ref{fig:modpo_potb} (left graph), the CR and OLR move to stronger perturbing amplitudes, which is partly the reason for the change of response. With an observed bar potential and
observed structures similar experiments should help to estimate the pattern velocity
and the true positions of the resonances.


\section{\Epic~applied to NGC~1365}
\label{sec:1365}

\begin{figure*}
   \centering
\includegraphics[width=0.9\linewidth, trim=0mm 0mm 0mm 0mm]{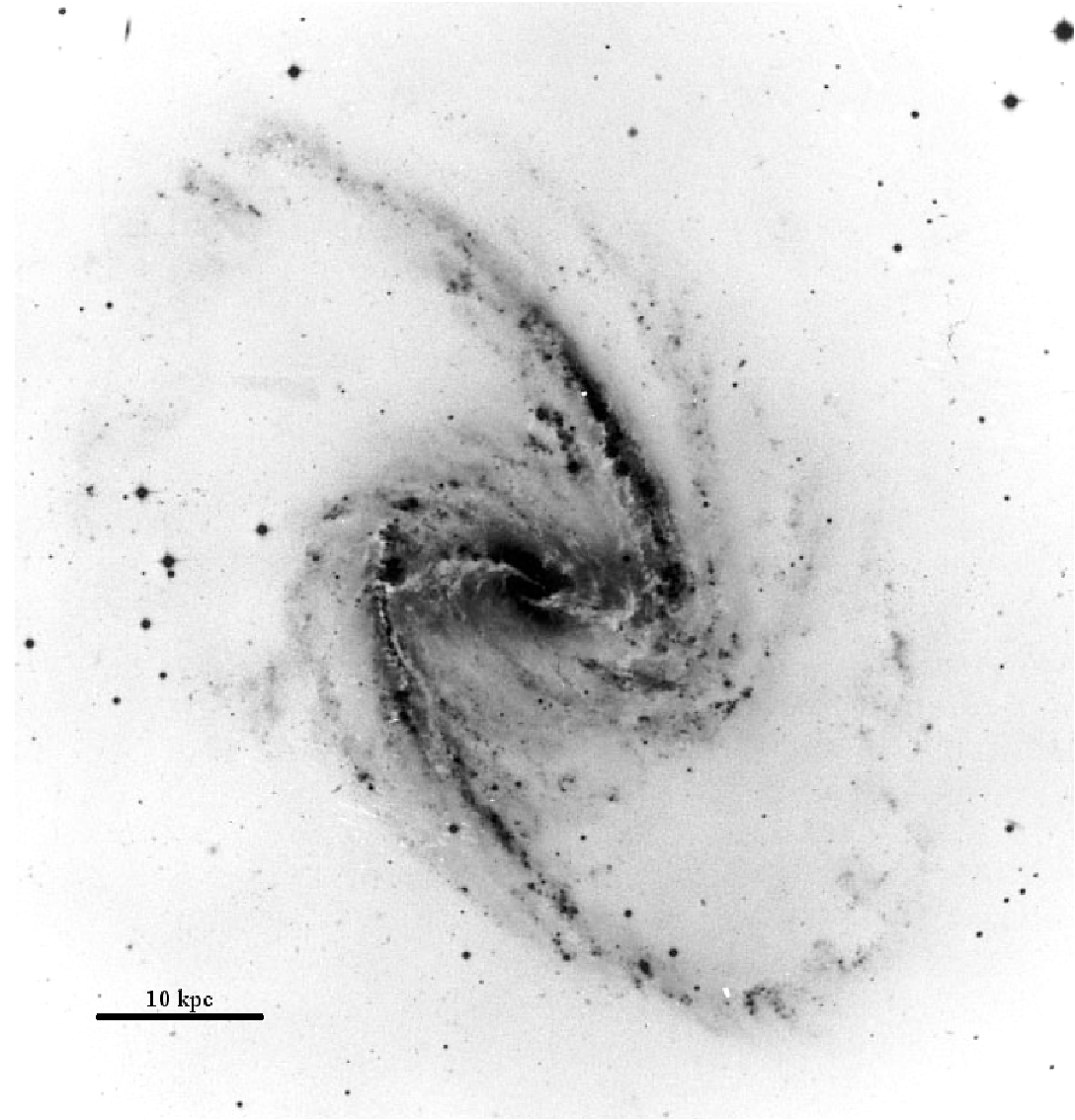}%
\caption{Optical image of NGC~1365 obtained at the prime focus of the ESO 3.6~m telescope.}
\label{fig:1365}
\end{figure*}

To evaluate the limitations and abilities of \Epic~we want to compare it to 
hydrodynamic simulations of a nearby barred galaxy. For this, the strongly 
barred Seyfert galaxy NGC~1365 (Fig. \ref{fig:1365}) would be suitable. This galaxy has 
been extensively 
observed (see the review by P.O. \citealt{Lindblad1999}) and a hydrodynamic simulation 
of the gas flow pattern was derived by LLA96. Also \citet{Halphangc1365} observed and modeled the gas flow in the bar. 

\begin{figure}
   \centering
\includegraphics[width=0.36\textwidth, trim=0mm 0mm 0mm 0mm, angle=-90]{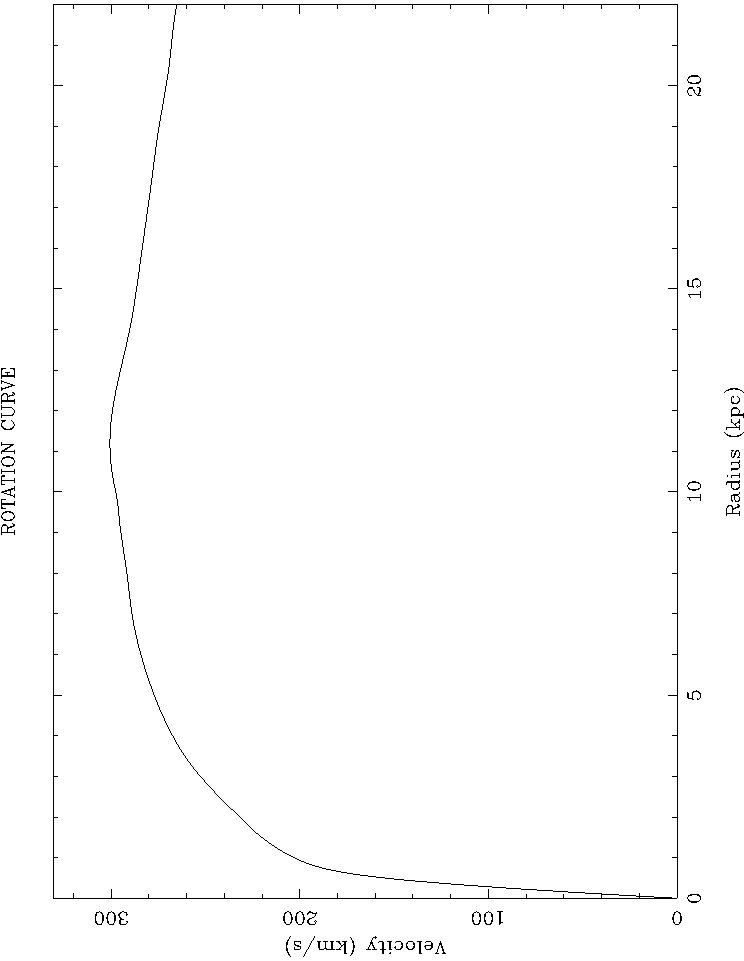}%
\caption{ Rotational velocity curve of NGC~1365 (from LLA96). }
\label{fig:input}
\end{figure}

A total hybrid velocity field has been derived by \citet{Lindbladpo1996} based on the HI velocity field of \citet{HIngc1365} and 
complemented with optical long slit spectra. We have here accepted the 
rotation curve based on this as derived by LLA96 (Fig. \ref{fig:input}). It agrees closely 
with the rotation curve given by \citet{Halphangc1365}.
%
%
We have derived the perturbed potential from the total surface density following the analysis using Bessel functions described in \citet[Ch. 2.6.2]{BinneyTremaine2008} and LLA96 (their Appendix A). For this analysis, we used the Fourier series decomposition of the total surface density, obtained by LLA96 from a J-band image as given in their Figs. 7 and 9, and a normalized triangle density distribution along the vertical direction, with a $z_0$ scale parameter equal to 1~kpc. 
%
%
We smoothed the rotational velocity curve and the Fourier components along the radius and interpolated the curves in order to achieve enough resolution for the $\lambda_2$ criterion of the spiral potential (eq. \ref{eq:condition}). We use a clock-wise rotation in order to reproduce the observed galaxy.


When comparing with observations, we have taken the results given by \Epic, located in the galactic plane, and transformed them into the sky plane. For this purpose, we have used a position angle, PA, equal to 40\degree~and an inclination angle, $i$, of 42\degree~ (\citealt{HIngc1365}, also used in LLA96).
In addition, we have assumed the systemic velocity to be 1632~\kms~and a scale of 0.1~kpc/arcsec.

\subsection{Comparing with observations and simulations}

The hydrodynamic simulations presented in LLA96 are performed using a 
flux-splitting 2nd-order code FS2, solving the flow of an ideal isothermal 
non-viscous gas \citep{vanAlbada1985}. LLA96 applies two different potentials 
derived from NGC~1365: bar only and bar + spiral. We have followed the same 
procedure in order to compare the results.

\subsubsection{Bar potential}

\begin{figure*}
   \centering
\includegraphics[width=0.48\linewidth, trim=0mm 0mm 0mm 0mm, angle=-90]{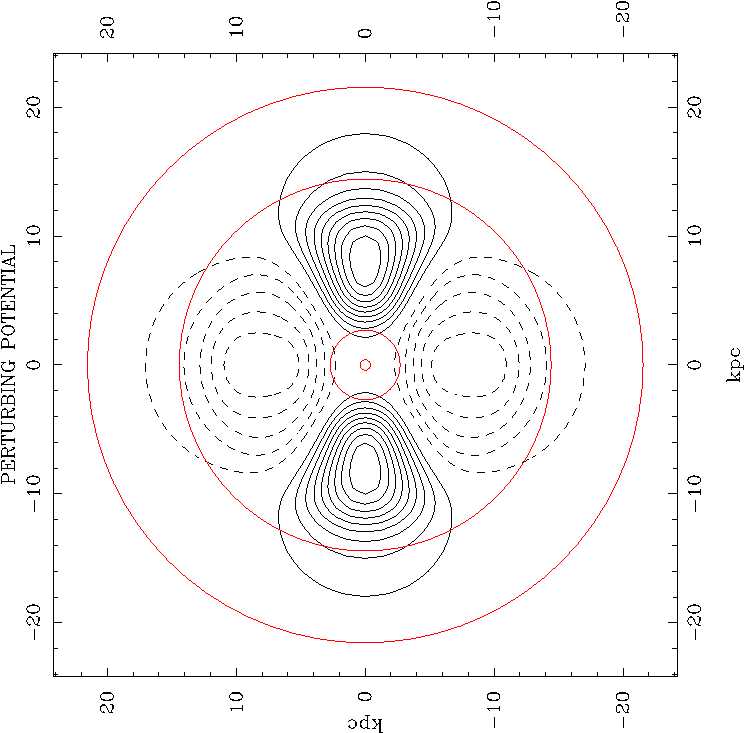}
\includegraphics[width=0.48\linewidth, trim=0mm 0mm 0mm 0mm, angle=-90]{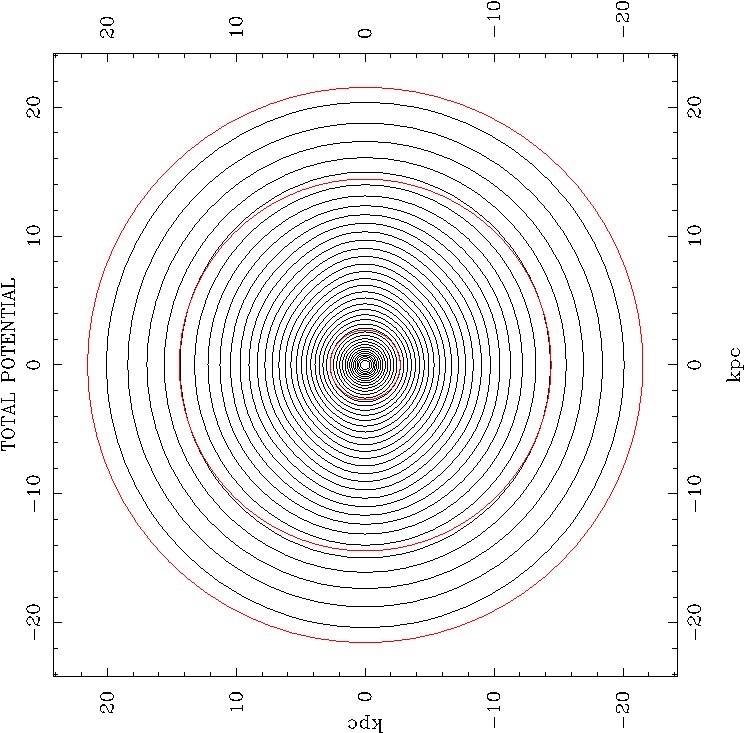}\\ \vspace{0.4cm}
\includegraphics[width=0.38\linewidth, trim=0mm 0mm 0mm 0mm, angle=-90]{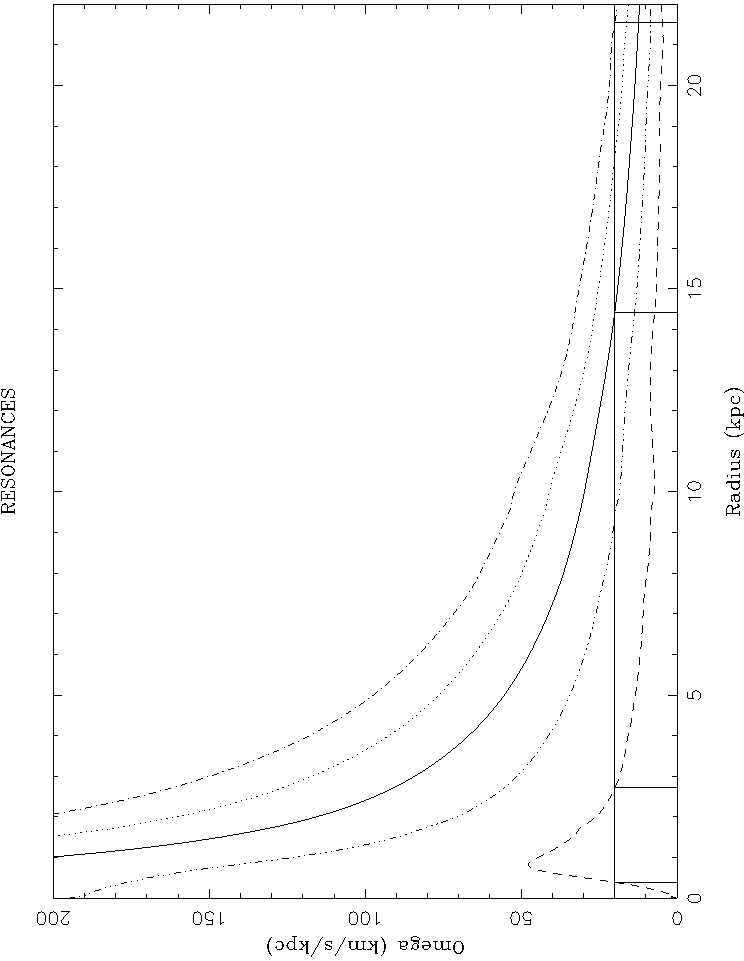}
\includegraphics[width=0.48\linewidth, trim=0mm 0mm 0mm 0mm, angle=-90]{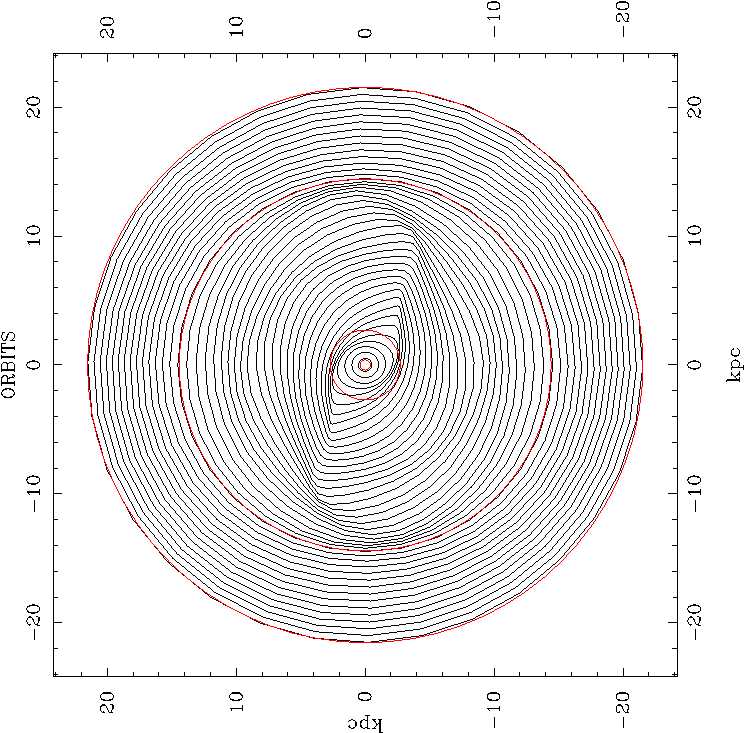}
\caption{\emph{Upper left graph:} Bar perturbing potential contours of NGC~1365, dashed lines correspond to positive contribution and solid lines to the negative contribution. \emph{Upper right graph:} Total potential contours of NGC~1365 for a bar strength, $|\Phi_1|_{\rm min}/|\Phi_0|_{\rm min}$, of 0.01. \emph{Red curves:} inner and outer Inner Lindblad Resonance (ILR), corotation and Outer Lindblad Resonance (OLR). \emph{Lower left graph:} Angular velocities as in Fig. \ref{fig:modpo_ome}. The adopted pattern speed of 20~\kmskpc (horizontal solid line) is indicated as well as the positions of the resonances. 
\emph{Lower right graph:} Orbits, seen as face-on, generated by \Epic~and the bar potential with a constant friction coefficient, $\lambda$, equal to 7~\kmskpc and a corotation damping coefficient,  $\varepsilon$, of 5 \kmskpc.  
}
\label{fig:inputbar}
\end{figure*}

When deriving the bar only potential, only the cosine contributions to the 
Fourier series has been considered, and only up to a radius of around 100~arcsec, where the phase shift of the different modes start a steep increase 
(see Figs. 8, 9 and 11 in LLA96). The bar is positioned horizontally along 
the x-axis (Fig. \ref{fig:inputbar}, upper left). 

     As in LLA96, we have used a pattern speed of 20 \kmskpc. As 
illustrated in Fig. \ref{fig:inputbar} (lower left), this 
pattern speed fixed the resonance radia to: the inner inner Lindblad resonance (iILR) at 0.4 kpc, the outer inner Lindblad resonance (oILR) at 2.7 kpc, the corotation radius (CR) at 14.4 kpc and the outer Lindblad resonance (OLR) at 21 kpc. 
We have used a constant friction coefficient $\lambda$ of 7~\kmskpc, and a 
corotation damping coefficient $\varepsilon$ of 5~\kmskpc. 
We are using an amplitude for the perturbing potential $\rm A_{\rm bar}=0.5$, half the amplitude used in LLA96. Thus, we have a potential with a bar strength, $|\Phi_1|_{\rm min}/|\Phi_0|_{\rm min}$, of 0.01. Increasing the bar amplitude, means that we have to increase the damping coefficients correspondingly to avoid self-crossing orbits, that are not realistic in a gaseous flow, and the results will be nearly the same. This may imply that the bar in NGC~1365 is actually too strong to be dealt with \Epic.

\begin{figure*}
   \centering
\includegraphics[width=0.33\linewidth, trim=10mm 0mm 40mm 0mm, angle=0]{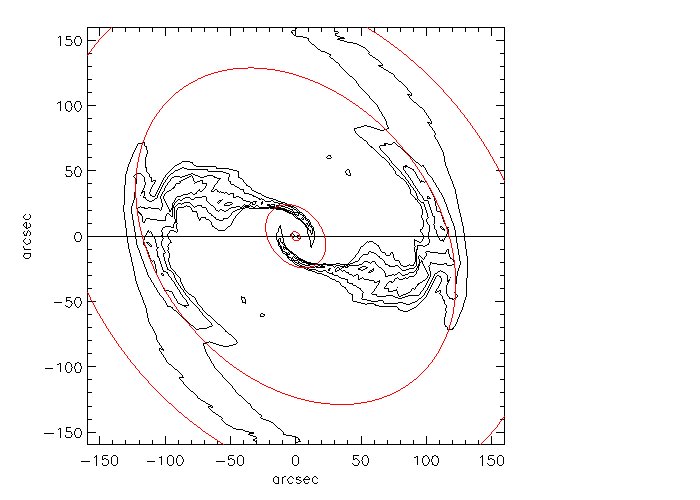}
\includegraphics[width=0.33\linewidth, trim=10mm 0mm 40mm 0mm, angle=0]{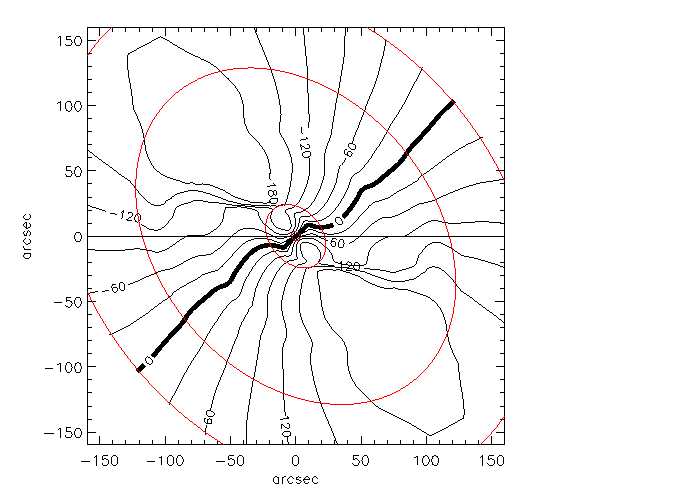}
\includegraphics[width=0.33\linewidth, trim=10mm 0mm 40mm 0mm, angle=0]{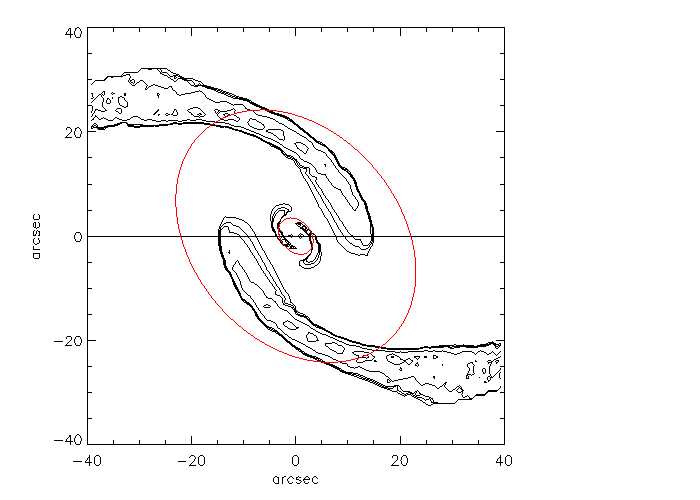}
\caption{
\emph{Left graph:} Density ratio, $\rho/\rho_0$. Contours at 1.06, 1.2, 1.4 and 1.6. \emph{Middle graph:} Line of sight velocity map derived by \Epic~using the position angle and the inclination of NGC~1365. Contours every 30 \kms, and thick solid line for the systematic velocity of the galaxy, 1632~\kms. \emph{Right graph:} Density ratio, $\rho/\rho_0$, for the inner 4~kpc. Contours at 1.02, 1.04, 1.06, 1.2, 1.4 and 1.6. Maps seen projected on the sky plane.
}
\label{fig:outputbar}
\end{figure*}

Using these input parameters, we obtained with \Epic~the orbits, densities and velocity field shown in Fig. \ref{fig:inputbar} and Fig. \ref{fig:outputbar}. The orbits are seen face-on. The density and velocity maps are projected on the sky plane, for comparison with the results of the simulations derived in LLA96 as well as with the observations. 

We see from Fig. \ref{fig:inputbar} (lower right) that the orbits in the region around oILR take 
elliptical shape, twisting in the counter rotation direction from being 
orientated at large angles to the bar, over 45~degrees at the oILR, to more 
elongated with the bar outside this resonance. Inside corotation matter 
circulates clockwise with respect to the bar, and the twisting and increasing 
deviation from elliptical shape gives rise to density increase forming  a 
small trailing spiral across the oILR. This straightens out to a lane along the 
leading edge of the bar with sharp shock-like velocity jumps over the lane. 

     The influence of the inner 1:4 resonance, that occurs around R = 9~kpc, 
can also be seen in the orbits. Even the 1:6 resonance further out is 
discernible in the slight crowding of orbits. Strong crowding of orbits are seen at CR at the Lagrangian points L1 and L2 at the end of the bar and, 
with slightly spiral form, at  the OLR in a direction perpendicular to the bar. 

     Compared to the orbits of the BM model of LLA96 (their Fig.~16), we see 
that between oILR and CR the BM orbits are more elongated and more orientated 
along the bar than ours. The 1:4 resonance is clearly seen also here. Around 
corotation the BM model displays the pendulum like motion referred to before 
and which is not reproduced by the linear approximation of \Epic. 

The interstellar matter density map (Fig. \ref{fig:outputbar} left graph), now projected on 
the sky plane, should be compared with the optical image of the galaxy (Fig. \ref{fig:1365}) 
and LLA96 (their Fig. 15). As expected from the orbits, the map shows a lane 
on the leading side of the bar that bends into spiral shape over the oILR. 
At the position of the 1:4 resonance a faint spur is seen, a strong arm 
appears at the end of the bar along the corotation radius, and a faint 
trailing arm extends out to the OLR. All these features have their 
correspondence in the BM density map of LLA96 (their Fig. 15). Fig. \ref{fig:outputbar}, right graph, shows 
the innermost region of the EPIC5 density map, where, as expected, a small 
nuclear leading spiral is seen across the iILR.

     The velocity map, to be compared with LLA96 (their Fig. 20) is shown 
in Fig. \ref{fig:outputbar} (middle graph). The map is projected on the sky plane with a line 
of nodes in PA 40\degree~and an inclination between the galaxy symmetry plane 
and the sky plane of $i$ = 42\degree. The main difference, comparing the \Epic~ 
result against the observations and simulations, is that the twisting of the 
zero velocity contour towards the bar is much weaker in the \Epic~case, which 
is connected with the different shapes of the orbits in the bar discussed 
above. This may partly be due to the difficulty of \Epic~ to accept a very 
strong bar, as in NGC~1365, without getting crossing orbits. The behavior 
of the velocity contours inside the oILR are similar to that in LLA96, 
and the sharp velocity jumps, of the order of 100~\kms, in the lanes on the leading side of the bar are 
seen in all cases. The twisting of the contours along the spiral arms with 
velocity jumps on the inner side of the arms inside corotation are seen 
both in the simulations and in the \Epic~results.

\subsubsection{Bar + Spiral potential}
\label{subsec:bar+spiral}

\begin{figure*}
   \centering
\includegraphics[width=0.48\linewidth, trim=0mm 0mm 0mm 0mm, angle=-90]{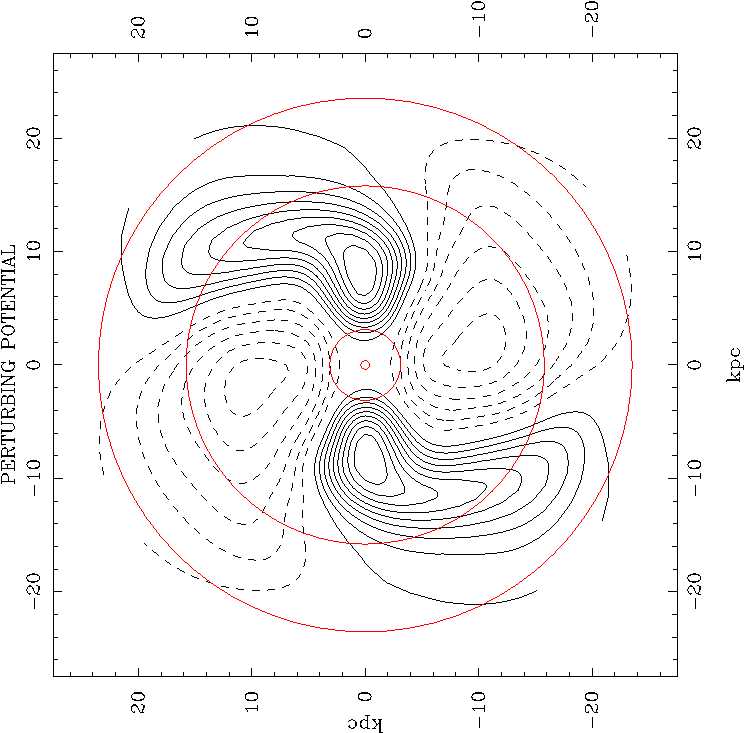}
\includegraphics[width=0.48\linewidth, trim=0mm 0mm 0mm 0mm, angle=-90]{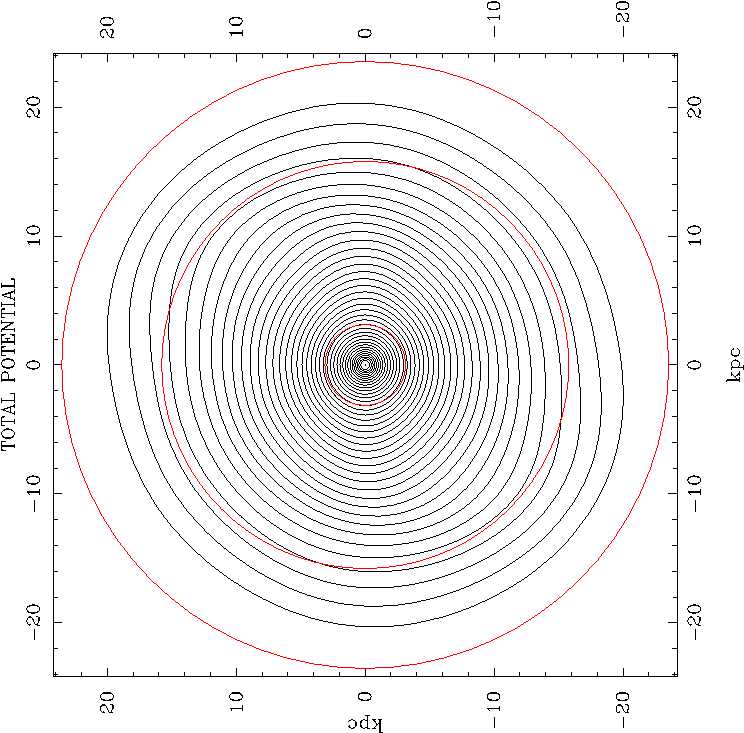}\\ \vspace{0.4cm}
\includegraphics[width=0.38\linewidth, trim=0mm 0mm 0mm 0mm, angle=-90]{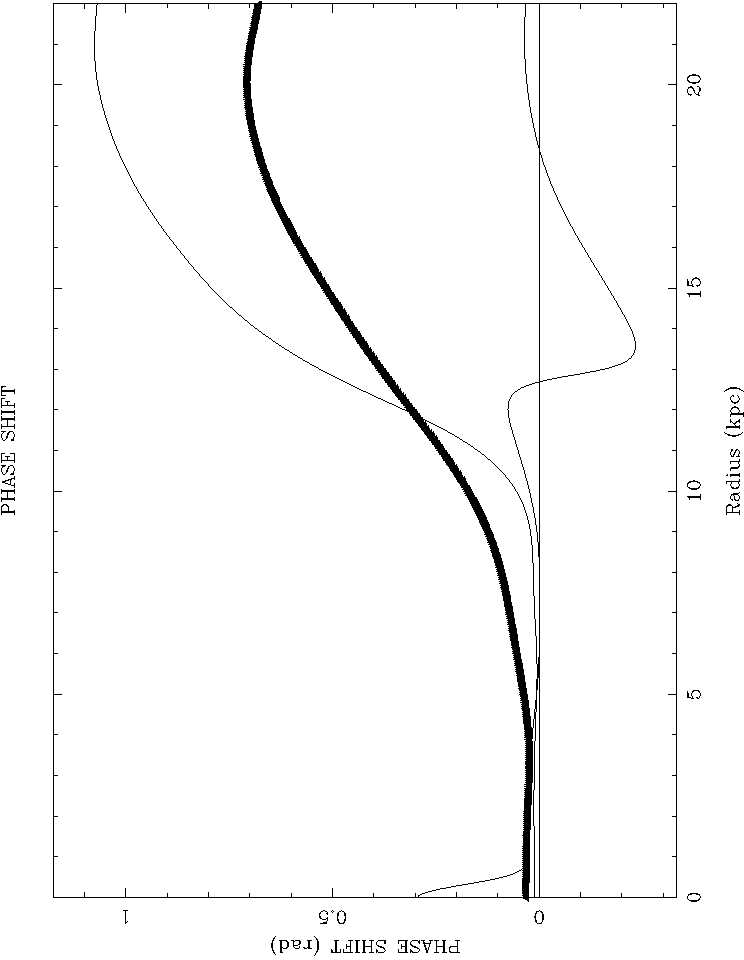}
\includegraphics[width=0.48\linewidth, trim=0mm 0mm 0mm 0mm, angle=-90]{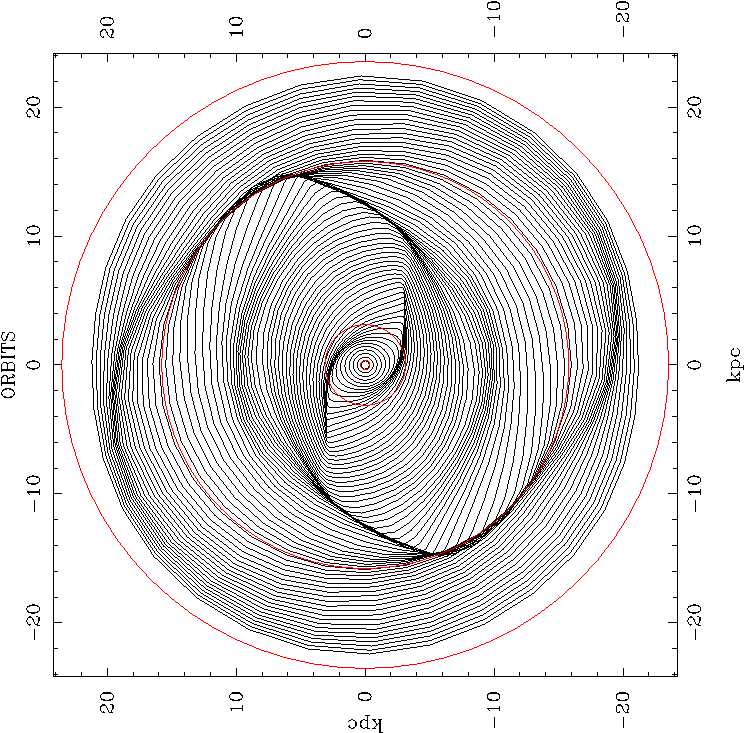}
\caption{\emph{Upper left graph:} Bar + spiral perturbing potential contours of NGC~1365. \emph{Upper rigth graph:} Total potential contours of NGC~1365 corresponding to a perturbing strength of the potential, $|\Phi_1|_{\rm min}/|\Phi_0|_{\rm min}$, equal to 0.01. \emph{Red curves:} inner and outer Inner Lindblad Resonance (ILR), corotation and Outer Lindblad Resonance (OLR). \emph{Lower left graph:} Phase shift of the perturbed potential, $\vartheta_m(r)$, presented in the appendix \ref{ap:theory}, for the three considered modes of its Fourier development. Thick solid line correspond to $m$ equal 2. \emph{Lower right graph:} Orbits seen as face-on generated by \Epic~using the bar + spiral potential and a constant friction coefficient, $\lambda$, equal to 7 \kmskpc ~and a corotation damping coefficient, $\varepsilon$, of 5 \kmskpc.}
\label{fig:inputspiral}
\end{figure*}

\begin{figure*}
   \centering
\includegraphics[width=0.48\linewidth, trim=10mm 0mm 40mm 0mm, angle=0]{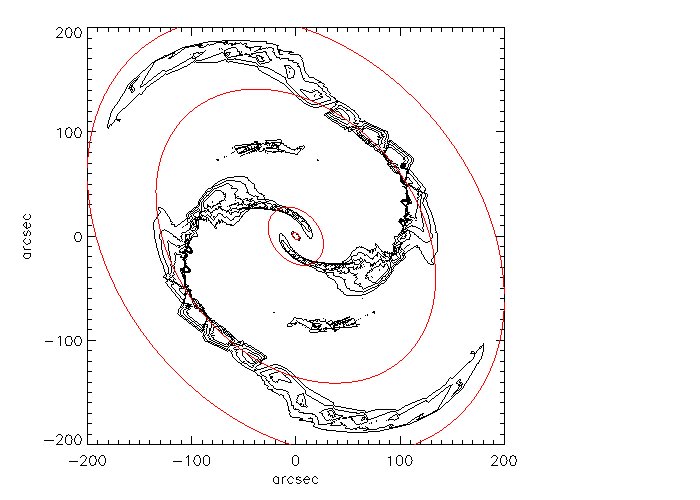}
\includegraphics[width=0.48\linewidth, trim=10mm 0mm 40mm 0mm, angle=0]{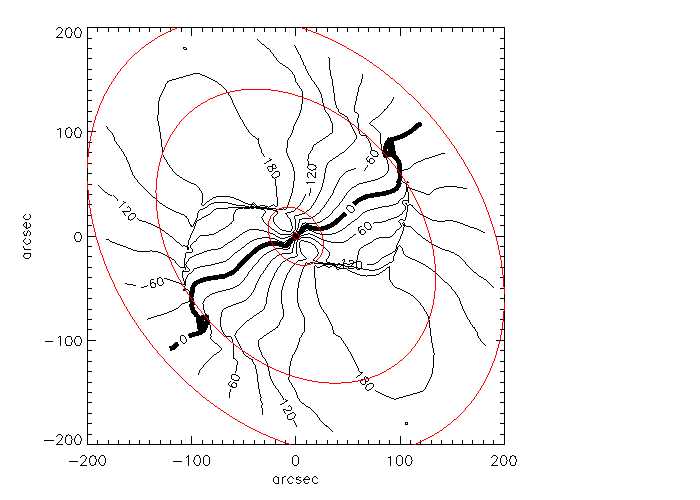} \\ \vspace{0.4cm}
\includegraphics[width=0.48\linewidth, trim=0mm -10mm -10mm 0mm, angle=0]{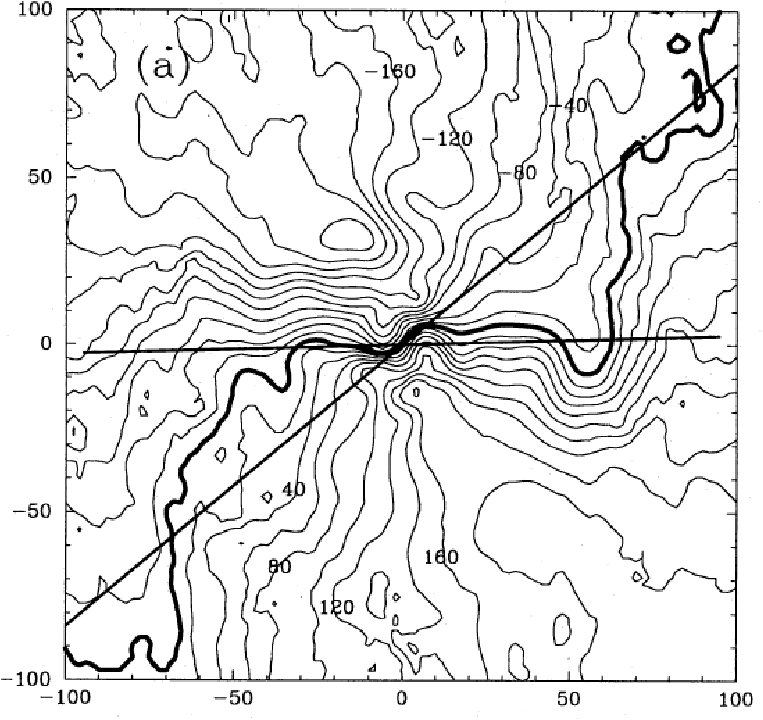}
\includegraphics[width=0.48\linewidth, trim=10mm 0mm 40mm 0mm, angle=0]{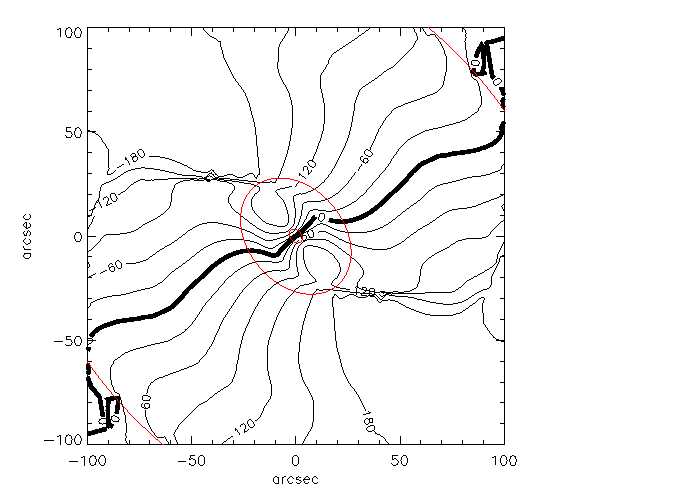}
\caption{\emph{Upper left graph:} Density ratio, $\rho/\rho_0$, for bar + spiral potential. Contours at 1.06, 1.2, 1.4 and 1.6. \emph{Upper right graph:} Line of sight velocity map derived by \Epic~ using the position angle and the inclination of NGC~1365. Contours every 30 \kms, and thick solid line for the systematic velocity of the galaxy, 1632 \kms. 
\emph{Lower left graph:} Optical radial velocity field in the inner 100~arcsec of NGC~1365 taken from LLA96. \emph{Lower right graph:} Line of sight velocity map derived by \Epic~for a bar + spiral potential in the inner 100~arcsec. 
Maps seen projected on the sky plane.
}
\label{fig:outputspiral}
\end{figure*}

In the case of the bar + spiral potential we have taken the sine and cosine contribution of the Fourier series of the first three even modes until a radius of around 220 arcsec, just as described in LLA96. The resulting potential is seen in Fig. \ref{fig:inputspiral}. We have used a pattern speed of 18 \kmskpc~(found for this potential by LLA96) and placed the bar horizontally along the \emph{x}-axis. This pattern speed makes the resonance radia to be: the inner inner Lindblad resonance (iILR) at 0.38 kpc, the outer inner Lindblad resonance (oILR) at 3.1 kpc, the corotation radius (CR) at 15.8 kpc and the outer Lindblad resonance (OLR) at 23.5 kpc. We have chosen the coefficients $\lambda$ and $\varepsilon$ and parameter $A_{bar}$ the 
same as in the bar only case. The main difference between this case and the 
bar only case is that the spiral features are led much easier through 
corotation, which can be seen both in the orbits, Fig. \ref{fig:inputspiral} lower right graph, and the densities, Fig. \ref{fig:outputspiral} upper left graph. This was 
noticed also by LLA96. 

\begin{figure*}
\centering
\includegraphics[width=0.48\linewidth, trim=15mm 0mm 15mm 0mm, angle=0]{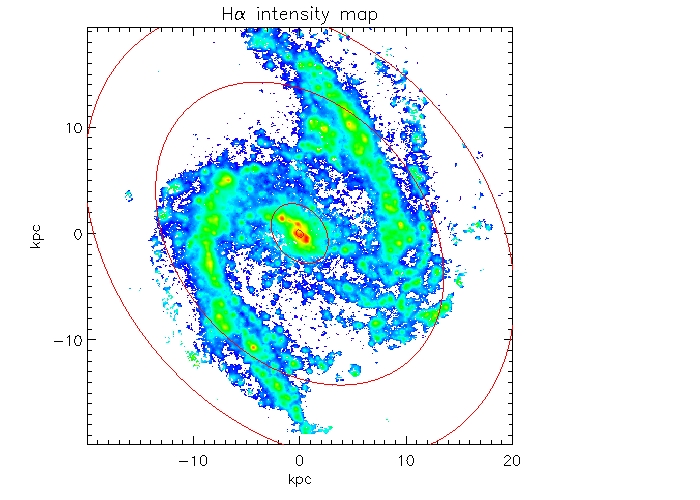}
\includegraphics[width=0.48\linewidth, trim=5mm 0mm 25mm 0mm, angle=0]{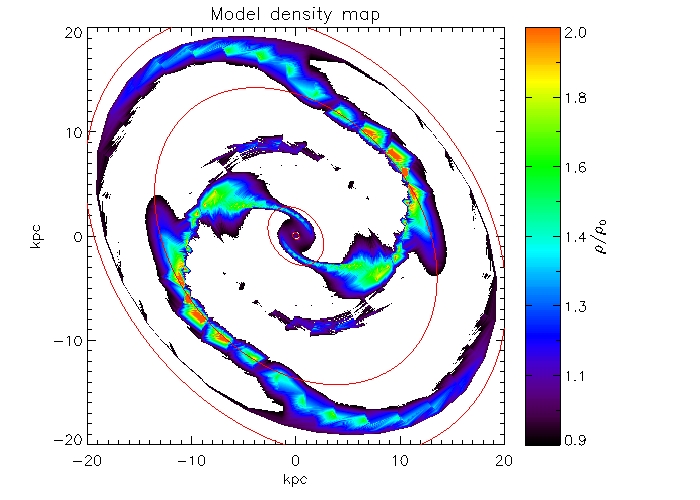}
\caption{\emph{Upper left map:} H$\alpha$ intensity map observed using a Fabry-Perot interferometer by \citet{Halphangc1365}. Shown in arbitrary units. \emph{Upper right map:} Model density ratio $\rho/\rho_0$ obtained by \Epic~, eq. (\ref{eq:density}). Ratios lower than 0.9 are in white color.
}
\label{fig:outputobs}
\end{figure*}

Fabry-Perot interferometry of NGC~1365, covering the H$\alpha$ regime has been
obtained by \citet{Halphangc1365}. By kind permission we have got 
access to their total H$\alpha$ intensity data, which are reproduced in Fig. \ref{fig:outputobs}  
(left map). In Fig. \ref{fig:outputobs} (right map) we have for comparison color coded the 
\Epic~ intensities from Fig. \ref{fig:outputspiral}. Clearly the \Epic~model agrees well with the 
H$\alpha$ structure. The hot spots in the nuclear region and the active Seyfert 
nucleus is of course not reproduced by \Epic.

\begin{figure*}
\centering
\includegraphics[width=0.48\linewidth, trim=20mm 0mm 10mm 0mm, angle=0]{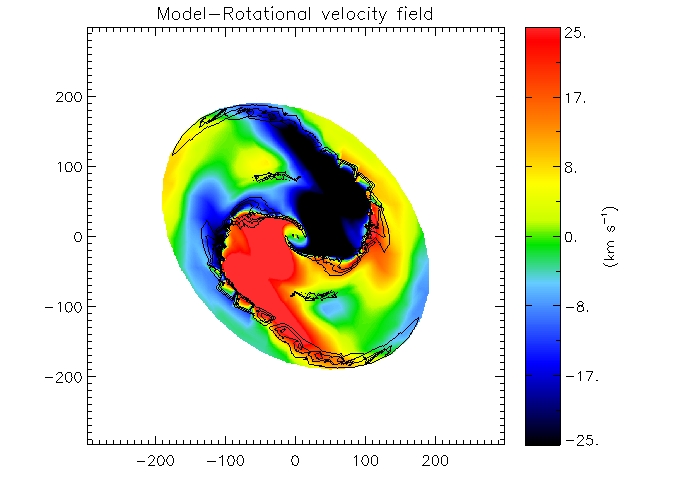} 
\includegraphics[width=0.4\linewidth, trim=0mm 4mm 20mm 0mm, angle=0]{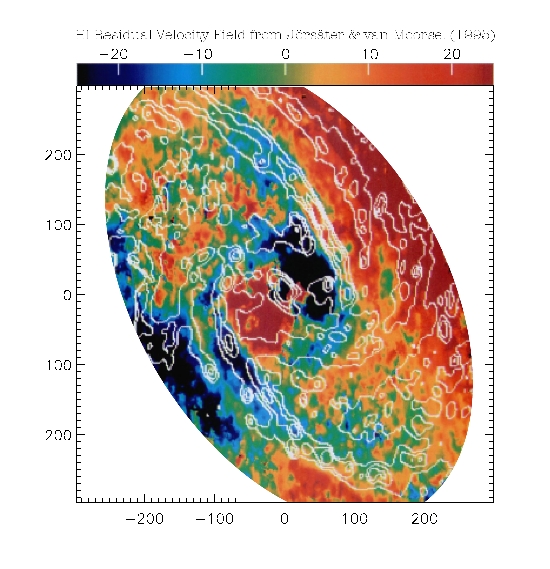}
\caption{\emph{Left graph:} Velocity model generated by \Epic~for a bar + spiral potential minus the rotational velocity from Fig. \ref{fig:input}. Density contour over-plot in solid black lines. \emph{Right graph:} HI velocity field observations minus rotational velocity model, figure from \citet{HIngc1365}. The contours are from an optical image of NGC~1365. The NW side of the galaxy is the near one, and the rotation is clockwise. $x$-axis and $y$-axis in both graphs in arcsec.
}
\label{fig:m-r}
\end{figure*}

The velocity field obtained with \Epic~is shown in Fig. \ref{fig:outputspiral}, together with the optical radial velocity field of NGC~1365 (inner 100~arcsec) taken from LLA96. 
In the innermost part, well inside the oILR, the contours indicate rapid rotation in nearly 
circular orbits. Between oILR and CR we again see how the isovelocity contours are concentrated 
to the leading edge of the bar. In the \Epic~ case the contours are squeezed to a shock-like jump 
of about 100 km/s. The optical map cannot make such a sharp jump because it is smoothed due to 
the limited spatial resolution given by the limited set of long slits on which it is based. 
Over corotation we see similar shock behavior on the inner side of the spiral arms.
The residual map obtained by subtracting the rotational velocity, Fig. \ref{fig:input}, from the \Epic~model velocity field is presented in Fig. \ref{fig:m-r}. In this Fig. \ref{fig:m-r} is also presented the residual velocities taken from \citet{HIngc1365}. They subtracted from the observed HI velocity map, the rotational velocity derived from their observations considering a warp disc model after a radius of 250~arcsec. The general behavior is similar in the two cases with shocks along the bar and spiral arms. 
It is informative to compare the \Epic~ map with the orbital map in Fig.~\ref{fig:inputspiral} (lower right) to 
envision how the orbital circulation results in the observed line of sight velocity map.

\section{Conclusions}
\label{sec:conclusions}

\Epic~cannot fully replace elaborate hydrodynamical simulations. \Epic~ solves 
the equations of motion in a solid body rotating, time 
independent galactic potential. This is done in the first order epicyclic 
approximation. \Epic~ computes the deformation of the initially 
circular orbits of the guiding center, the velocity fields, and the structure 
created. \Epic~ pictures a steady state created by a weak perturbing potential, 
it does not give the time evolution of galactic structure. 
Due to the degeneracy of the strength of the gravitational potential and damping coefficients, we can only derive relative values of the density.

     Our comparison with a more elaborate simulation, as well as observations,
show that it still can reproduce structural and dynamic properties. One of 
the advantages with \Epic, involving an analytic solution, is the speed by 
which it can survey the parameter space. Further, the possibilities, by varying  
parameters, to isolate artificially various agents important for forming 
structure of galaxies will be of importance for understanding the formation of 
galactic structure. \Epic~will be a useful complement to time consuming, 
elaborate simulations.

\vspace{5mm}

We would like to thank to J. A. Sellwood and R. Z\'anmar S\'anchez for letting us use their H$\alpha$ data of NGC~1365. We also want to thank to the reviewer, Gene Byrd, for his useful comments. NP-F acknowledges financial support from NOTSA, and KF is supported by the Swedish Research Council (Vetenskapsr\aa det). 

\bibliographystyle{mn2e}
\bibliography{mybib}

\onecolumn

\appendix
\section{Perturbations in the epicyclic approximation}
\label{ap:theory}
\indent
We present in this paper a solution of the motion of a mass particle in an axisymmetric potential, $\Phi_0(r)$, with a weak non-axisymmetric perturbation, $\Phi_1(r,\theta)$. We assume that the potential is rotating with a pattern speed \Omegap, and we study the system in a corotating frame. We express the potential in polar coordinates:

\begin{equation}
\Phi(r,\theta)=\Phi_0(r)+\Phi_1(r,\theta)=\Phi_0(r) - \sum_{m=1}^{n} \Psi_m(r) \, \cos m(\theta - \vartheta_m(r))
\label{eq:potentialap}
\end{equation}

\noindent
where $\vartheta_m$ is the spiral phase. The rotation of the axisymmetric potential is about the center of gravity, which we have assumed to be at rest. Thus one has to be cautious in the presence of strong $m=1$ perturbation, when the center of mass could be significantly displaced.

In a corotating coordinate system we can write the equations of motions (see also \citealt{BinneyTremaine2008}, p.189):

\begin{align}
\ddot{r}-r\dot{\theta}^2&=-\frac{\partial \Phi}{\partial r} + 2 r \dot{\theta}\Omega_p + r \Omega_p^2 \\
r\ddot{\theta}+2\dot{r}\dot{\theta}&=-\frac{1}{r}\frac{\partial \Phi}{\partial \theta}-2\dot{r}\Omega_p
\end{align}

We introduce $\xi(t)$ and $\eta(t)$ as deviations from circular motion and write:


\begin{align}
\centering
r&=r_0+\xi \\
\theta&=\theta_0 + (\Omega - \Omegap) \,t + \frac{1}{r_0} \eta
\end{align}

\noindent
where $\Omega$ is the circular angular velocity at a radius $r_0$. Assuming $\xi$ and $\eta$ to be small, and linearizing the equations of motion by neglecting higher order terms of $\xi$ and $\eta$, we get:

\begin{align}
\centering
\ddot{\xi} - 2 \Omega\dot{\eta} - 4 \Omega A \xi &= - \frac{\partial \Phi_1}{\partial r} \label{eq:eqofmotion1}\\
\ddot{\eta} + 2 \Omega \dot{\xi} & = - \frac{1}{r} \frac{\partial \Phi_1}{\partial \theta}
\label{eq:eqofmotion}
\end{align}
\noindent
where $A$ is the Oort constant, $A=-r/2 \,\,\, d\Omega/dr$.

To describe the motion of a gaseous medium, we introduce a frictional force proportional to the deviation from circular motion with a damping coefficient, $2\lambda$ (see \citealt{Wada1994}, \citealt{Lindblad1994}). With the introduction of this frictional force 
%
the equations of motion, eq. (\ref{eq:eqofmotion1}) and (\ref{eq:eqofmotion}), are transformed into the equations (\ref{eq:eqofmotionfric}). 

\begin{align}
\centering
\ddot{\xi} + 2 \lambda \dot{\xi} - 2\Omega \dot{\eta} - 4 \Omega A \xi  &= - \frac{\partial \Phi_1}{\partial r} = 
\sum_{m=1}^{n} \big[ C_m \cos m(\theta - \vartheta_m) + E_m \sin m(\theta - \vartheta_m)\big] \nonumber\\
\ddot{\eta} + 2 \Omega \dot{\xi} + 2 \lambda \dot{\eta} + 4 \lambda A \xi &= - \frac{1}{r}\frac{\partial \Phi_1}{\partial \theta} =
- \sum_{m=1}^{n} D_m \sin m(\theta - \vartheta_m) \label{eq:eqofmotionfric}
\end{align}

\noindent
where

\begin{align}
\centering
C_m = \frac{d\Psi_m}{dr}; \;\;\;\;\; &D_m=m\frac{\Psi_m}{r}; \;\;\;\;\;  E_m=m\Psi_m\frac{d\vartheta_m}{dr} \nonumber
\end{align}

\noindent
Provided that we are not close to corotation, where $\Omega=\Omega_p$, we can assume that $\eta/r_0<<(\Omega-\Omega_p)$, what makes 

\begin{equation}
\theta\sim\theta_0 + (\Omega - \Omegap) \,t
\label{eq:aprox}
\end{equation}

\noindent
with $\kappa^2=4\Omega^2-4\Omega A$, we can write the full solution of the now linearized eqs. (\ref{eq:eqofmotionfric}) as:

\begin{align}
\xi &= c e^{-\alpha\lambda t} \cos \kappa(t-t_0) + \sum_{m=1}^{n} \big[ d_m \cos m(\theta-\vartheta_m) + e_m \sin m(\theta -\vartheta_m) \big] \label{eq:xisolap}\\
\eta&= - \frac{2\Omega}{\kappa}c e^{-\beta\lambda t} \sin \kappa(t-t_0) + \sum_{m=1}^{n} \big[ g_m \sin m(\theta-\vartheta_m) + f_m \cos m(\theta -\vartheta_m) \big]\label{eq:etasolap}
\end{align}

\noindent
$\alpha$ and $\beta$ are functions of $\Omega$ and $\kappa$, and $c$ is an arbitrary constant. The second terms on the left 
side of eqs. (\ref{eq:xisolap}) and (\ref{eq:etasolap}) give the forced oscillation due to the perturbing 
force. The first terms give the damped oscillation with the epicyclic frequency 
$\kappa$ around these guiding centra. We will leave out these latter terms in 
what follows and just consider the motions of the guiding center.

We introduce these eq. (\ref{eq:xisolap}) and eq. (\ref{eq:etasolap}) into (\ref{eq:eqofmotionfric}), getting a system of equations with the 4$n$ unknown amplitudes, $d_m$, $e_m$, $g_m$ and $f_m$:

\begin{align}
-(\omega_m^2 + 4 \Omega A ) d_m - 2 \omega_m \Omega g_m &= C_m -2\lambda \omega_m e_m \nonumber\\
-(\omega_m^2 + 4 \Omega A ) e_m + 2 \omega_m \Omega f_m &= E_m +2\lambda \omega_m d_m \nonumber\\
2\omega_m\Omega d_m + \omega_m^2 g_m &= D_m - 2\lambda \omega_m f_m + 4 \lambda A e_m \nonumber\\
2\omega_m\Omega e_m - \omega_m^2 f_m &= - 2\lambda \omega_m g_m - 4 \lambda A d_m
\label{eq:system}
\end{align}

\noindent
where $\omega_m = m\,(\Omega-\Omegap)$. The final solution is shown below.

\begin{align}
d_m&=\frac{(\kappa^2-\omega_m^2)(\omega_m C_m + 2 \Omega D_m) - 2 \lambda (\kappa^2+\omega_m^2)E_m - 4\lambda^2 (\omega_m C_m -2\Omega D_m) - 8\lambda^3 E_m}{\omega_\varepsilon \left[ \left(\kappa^2-\omega_m^2\right)^2 + 8\lambda^2\left(\kappa^2+\omega_m^2\right) + 16 \lambda^4 \right]} \label{eq:d} \\
 \nonumber\\
g_m&=\frac{-(\kappa^2-\omega_m^2)\left[2\Omega\left(\omega_m C_m +2\Omega D_m \right) - \left(\kappa^2 - \omega_m^2\right) D_m\right] + 4 \lambda \left[\left(\kappa^2-\omega_m^2\right)A+2\omega_m^2\Omega\right]E_m  }{\omega_\varepsilon^2 \left[ \left(\kappa^2-\omega_m^2\right)^2 + 8\lambda^2\left(\kappa^2+\omega_m^2\right) + 16 \lambda^4 \right]}  \nonumber \\
&\;+\frac{- 4 \lambda^2\left[\left(\omega_m C_m + 2 \Omega D_m \right) \left(2\Omega - 4 A\right) - \left(\omega_m^2+3\kappa^2-8\Omega^2\right)D_m\right] + 16 \lambda^3 A E_m}{\omega_\varepsilon^2 \left[ \left(\kappa^2-\omega_m^2\right)^2 + 8\lambda^2\left(\kappa^2+\omega_m^2\right) + 16 \lambda^4 \right]} \label{eq:g}\\ 
\nonumber\\
e_m &=\frac{\omega_m^2\left(\kappa^2-\omega_m^2\right)E_m + 2 \lambda \left[ \left( \kappa^2 + \omega_m^2\right) \left( \omega_m C_m + 2 \Omega D_m  \right) - 2 \Omega \left(\kappa^2-\omega_m^2 \right) D_m \right] - 4 \lambda^2 \omega_m^2 E_m + 8 \lambda^3 \omega_m C_m}{\omega_\varepsilon^2 \left[ \left(\kappa^2-\omega_m^2\right)^2 + 8\lambda^2\left(\kappa^2+\omega_m^2\right) + 16 \lambda^4 \right]} \label{eq:e} \\ \nonumber\\
f_m&=\frac{2\omega_m\left(\kappa^2-\omega_m^2\right)\Omega E_m + 2 \lambda \left[4 \omega_m \Omega \left( \omega_m C_m + 2 \Omega D_m\right) + \left(\kappa^2-\omega_m^2\right)\left(2AC_m - \omega_m D_m\right)\right]+8\lambda^2 \omega_m E_m \left(\Omega-2A\right) + 8 \lambda^3 \left(2AC_m+\omega_m D_m\right)
}{\omega_\varepsilon^2 \left[ \left(\kappa^2-\omega_m^2\right)^2 + 8\lambda^2\left(\kappa^2+\omega_m^2\right) + 16 \lambda^4 \right]} \label{eq:f}
\end{align}

\noindent
At the Lindblad resonances $\kappa^2-\omega_m^2=0$, and we can see how the friction terms are damping the resonances when $\lambda\neq 0$. At corotation $\omega_m=0$. To damp this resonance, as explained in Section \ref{sec:method} in the paper, we have replaced $\omega_m$ in the denominator of eqs. (\ref{eq:d}) to (\ref{eq:f}) by $\omega_\varepsilon$, where 
\begin{align}
\frac{1}{\omega_{\varepsilon}}=\frac{\omega_m}{\omega_m^2+\varepsilon_m^2}
\end{align}
where $\varepsilon_m = m\varepsilon$ and $\varepsilon$ is an additional parameter.

\end{document}